\documentclass[aps,prc,twocolumn,showpacs,floatfix,nofootinbib,preprintnumbers,superscriptaddress,amsmath,amssymb,longbibliography]{revtex4-1}

\usepackage{CJKutf8}
\usepackage[utf8]{inputenc}
\usepackage{amsmath} 
\usepackage{graphicx} 
\usepackage{hyperref}
\hypersetup{breaklinks=true,colorlinks=true,linkcolor=blue,citecolor=blue,filecolor=magenta,urlcolor=cyan}
\usepackage{amssymb}
\usepackage[mathscr]{euscript}
\renewcommand{\vec}{\boldsymbol}
\usepackage{xcolor}

\newcommand{\spin}{\textstyle\!\!\frac{1}{2}}
\newcommand{\smy}{{\mathfrak  r}_y}
\newcommand{\sms}{\breve{\sigma}_y}

\makeatletter
\def\@bibdataout@aps{%
\immediate\write\@bibdataout{%
@CONTROL{%
apsrev41Control%
\longbibliography@sw{%
    ,author="08",editor="1",pages="1",title="0",year="1"%
    }{%
    ,author="08",editor="1",pages="1",title="",year="1"%
    }%
  }%
}%
\if@filesw \immediate \write \@auxout {\string \citation {apsrev41Control}}\fi 
}
\makeatother

\begin{document}
\begin{CJK*}{UTF8}{gbsn}
\title{Nucleon localization function in rotating nuclei}

\author{T. Li (李通)}
\affiliation{National Superconducting Cyclotron Laboratory, Michigan State University, East Lansing, Michigan 48824, USA}
\affiliation{Department of Physics and Astronomy, Michigan State University, East Lansing, Michigan 48824, USA}

\author{M. Z. Chen (陈孟之)}
\affiliation{National Superconducting Cyclotron Laboratory, Michigan State University, East Lansing, Michigan 48824, USA}
\affiliation{Department of Physics and Astronomy, Michigan State University, East Lansing, Michigan 48824, USA}

\author{C. L. Zhang (张春莉)}
\affiliation{National Superconducting Cyclotron Laboratory, Michigan State University, East Lansing, Michigan 48824, USA}
\affiliation{Department of Physics and Astronomy, Michigan State University, East Lansing, Michigan 48824, USA}

\author{W. Nazarewicz}
\affiliation{Department of Physics and Astronomy, Michigan State University, East Lansing, Michigan 48824, USA}
\affiliation{Facility for Rare Isotope Beams, Michigan State University, East Lansing, Michigan 48824, USA}

\author{M. Kortelainen}
\affiliation{Department of Physics, PO Box 35 (YFL), FI-40014 University of
	Jyv\"{a}skyl\"{a}, Finland}
\affiliation{Helsinki Institute of Physics, P.O. Box 64, FI-00014 University of
	Helsinki, Finland}

\date{\today}

\begin{abstract}

\begin{description}
\item[Background]
An electron localization function was originally introduced  to visualize  in positional space bond structures in molecules. It became a useful tool to describe electron configurations in  atoms,  molecules, and solids. In nuclear physics,  a nucleon localization function (NLF) has been used  to characterize cluster structures  in light nuclei,  formation of fragments in fission, and pasta phases appearing in the inner crust of neutron stars. 

\item[Purpose]
We use  the NLF to study the nuclear response to fast rotation.

\item[Methods]
We generalize the NLF to the case of nuclear rotation. The extended expressions involve both time-even and time-odd local particle and spin densities and currents. Since 
the current density and  density gradient  contribute  to
the NLF primarily at the surface, we
 propose a simpler spatial  measure given by the kinetic energy density.
Illustrative calculations for the superdeformed yrast band of $^{152}$Dy were carried out by using the  cranked Skyrme-Hartree-Fock method. We also employed the cranked harmonic-oscillator model to gain insights into spatial
patterns revealed by the NLF at high angular momentum.

\item[Results]
In the case of a deformed rotating nucleus, several NLFs can be introduced, depending on the definition of the spin-quantization axis, direction of the total angular momentum, and self-consistent symmetries of the system.
Contributions to the NLF from the current density, spin-current tensor density,  and density gradient terms are negligible in the nuclear interior. The oscillating pattern of the simplified NLF
can be explained in terms  of a constructive interference  between  kinetic-energy and  particle densities.
The characteristic nodal pattern seen in the NLF in the direction  of the major axis of  a  rotating nucleus comes from  single-particle orbits carrying large aligned angular momentum. The  variation  of the NLF along the minor axis of the nucleus can be  traced back to deformation-aligned orbits.

\item[Conclusions]
The NLF  allows a simple interpretation  of  the shell structure evolution in the rotating nucleus in terms of the angular-momentum alignment of individual nucleons. 
We expect that the NLF will be very useful for the characterization and visualization  of other collective modes in nuclei and time-dependent processes. 

\end{description}
\end{abstract}
\maketitle
\end{CJK*}

\section{Introduction}

Nuclear collective motion, such as rotations and vibrations, provides rich information about nuclear  structure and nuclear response to external fields.
When discussing nuclear collective motion,
one is often making analogies to molecules and their collective modes.  One has to bear in mind, however, that  the  $A$-body nuclear wave function cannot, in general,
be expressed in terms of slow and fast components because the time separation between single-particle (s.p.)  and collective nuclear motion is poor. Consequently,
deviations
from the  perfect rotational and vibrational patterns are abundant. Such deviations indicate that the nuclear collective modes
result from  coherent superpositions of individual  nucleonic excitations.

The observation of rotational bands in atomic nuclei has provided us with many insights into nuclear deformations and the underlying  shell structure \cite{Boh75,szymanskiFastNuclearRotation1983,Nazarewicz2001,frauendorfSpontaneousSymmetryBreaking2001}. 
Theoretically, high-spin states can be described in a fully self-consistent way by 
the nuclear energy density functional (EDF) method   \cite{benderSelfconsistentMeanfieldModels2003},
which is  closely related to density-functional theory \cite{Duguet2010,Drut2010}.
Although rotation is essentially a time-dependent problem, the introduction of a rotating intrinsic frame through the cranking approximation transforms the time-dependent   problem into a time-independent one \cite{Nakatsukasa2016}.
The cranking term added to the nuclear Hamiltonian can  be interpreted as a constraint on the angular momentum, with the rotational frequency  playing the role of  the Lagrange multiplier. 

The  spatial electron localization function (ELF)
was originally introduced in the context of electronic Hartree-Fock (HF) studies  to characterize shell structure in atoms and chemical bonds in molecules 
\cite{beckeSimpleMeasureElectron1990, savinElectronLocalizationSolidState1992, savinELFElectronLocalization1997,  burnusTimedependentElectronLocalization2005, fuentealbaChapterUnderstandingUsing2007,jerabekElectronNucleonLocalization2018}. 
In nuclear structure research, the nucleon localization function (NLF)
 turned out to be a useful tool for the identification of clusters in light nuclei \cite{reinhardLocalizationLightNuclei2011,Schuetrumpf2017a,Ebran2017} and nuclear reactions \cite{schuetrumpfClusterFormationPrecompound2017}; formation of fragments in fission
\cite{zhangNucleonLocalizationFragment2016,Scamps2018,Scamps2019,Matheson2019,Sadhukhan2020}; and
nuclear pasta phases in the inner crust of neutron stars \cite{Schuetrumpf2017a}. 
Compared with nucleonic distributions that are fairly constant in the nuclear interior, the NLF more effectively quantifies nuclear configurations through its characteristic oscillating pattern due to shell effects. Consequently, it is expected to be a good indicator of the competition between s.p.\ motion and collective nuclear modes.

In this work,  we  use the NLF  to study the  nuclear response to rotation.
We consider the case of superdeformed (SD)  $^{152}$Dy, a  quintessential  nuclear rotor  that has been investigated in a number of self-consistent works \cite{dobaczewskiTimeoddComponentsMean1995, satulaAdditivityQuadrupoleMoments1996,afanasjevTimeoddMeanFields2010}. 

This paper is organized as follows:
Our theoretical framework is described in  Sec. \ref{sec:theory}, which contains    a comprehensive discussion of the NFL and its extension to the case of rotation. 
Section.~\ref{sec:results} contains the results of HF calculations for $^{152}$Dy that are supplemented by cranked harmonic-oscillator model results that illuminate essential points. 
Finally, Sec.~\ref{sec:conclusion}  presents conclusions
and perspectives for future studies.

\section{Theoretical framework}\label{sec:theory}

\subsection{Density matrices}\label{sec:dens}

The starting point in the derivation of the spatial localization function is 
 the one-body HF density matrix in the coordinate representation:
 \begin{equation}\label{nldens}
  \rho(\vec{r}s q,\vec{r}'s' q')
  \equiv\langle \Psi|a^{\dagger}_{\vec{r}'s' q'}a_{\vec{r}s q}|\Psi \rangle,
\end{equation}
where $a^{\dagger}_{\vec{r}s q}$ and $a_{\vec{r}s q}$
create and annihilate, respectively, a nucleon $q$ (=$n$ or $p$)
at point $\vec{r}$ with spin $s=\pm\,\,\spin$, and   $|\Psi \rangle$  is the HF independent-particle state. 
In what follows, we consider pure proton and neutron HF
states, i.e., $q'=q$, and we define $\rho_q(\vec{r} s, \vec{r}' s' )=\rho(\vec{r}s q,\vec{r}'s' q)$. 

Expressed in terms of spin components, the nonlocal HF density matrices can be written as \cite{engelTimedependentHartreefockTheory1975,benderSelfconsistentMeanfieldModels2003,Perlinska2004}:
\begin{equation}\label{eq:Engle_start_nonlocal}
\rho_q(\vec{r} s, \vec{r}' s' ) =  \frac{1}{2}\left[ \rho_q(\vec{r}, \vec{r}') \delta_{ss'} 
 + (s|\vec{\sigma}| s') \vec{s}_q(\vec{r}, \vec{r}')\right],
\end{equation}
where
\begin{subequations}
\begin{align}
\rho_q(\boldsymbol{r}, \boldsymbol{r}') &=\sum_{s} \rho_q\left(\boldsymbol{r} s, \boldsymbol{r}' s\right), \\
\boldsymbol{s}_q\left(\boldsymbol{r}, \boldsymbol{r}'\right) &=\sum_{ss'} \rho_q\left(\boldsymbol{r} s, \boldsymbol{r}' s'\right) 
\langle s'|\boldsymbol{\sigma}| s\rangle.
\end{align}
\end{subequations}

In the EDF method with the zero-range Skyrme interaction, the energy functional depends only on local densities and currents. Following the
standard definitions \cite{engelTimedependentHartreefockTheory1975,benderSelfconsistentMeanfieldModels2003}, in the present study we employ the following  densities:
\begin{subequations}\label{densitiesall}
\begin{align}
\rho_q(\vec{r}) &= \rho_q(\vec{r},\vec{r}), \label{rhodens} \\
\vec{s}_q(\vec{r}) &= \vec{s}_q(\vec{r},\vec{r}), \label{sdens} \\
\tau_q(\vec{r}) &= \left[\vec{\nabla}\cdot\vec{\nabla}^\prime\rho_q(\vec{r},\vec{r}^\prime)\right]_{\vec{r}=\vec{r}^\prime}, \label{kindens}  \\
\vec{j}_q(\vec{r}) &= \frac{1}{2i} \left[\left(\vec{\nabla}-\vec{\nabla}^\prime\right)\rho_q(\vec{r},\vec{r}^\prime)\right]_{\vec{r}=\vec{r}^\prime}, \label{curdens}  \\
\mathbb{J}_q(\vec{r}) &= \frac{1}{2i} \left[\left(\vec{\nabla}-\vec{\nabla}^\prime\right)\otimes\vec{s}_q(\vec{r},\vec{r}^\prime)\right]_{\vec{r}=\vec{r}^\prime}, \label{spincurdens}  \\
\vec{T}_q(\vec{r}) &= \left[(\vec{\nabla}\cdot\vec{\nabla}^\prime)\vec{s}_q(\vec{r},\vec{r}^\prime)\right]_{\vec{r}=\vec{r}^\prime} \label{spinkindens} ,
\end{align}
\end{subequations}
where
$\otimes$ stands for the tensor product of vectors in the physical space.

\subsection{Nucleon localization function}\label{sec:NLF}

Let us first consider the probability of finding two nucleons of  a given 
isospin $q$  and spin  $s$ at spatial
locations $\vec{r}$ and $\vec{r'}$:
\begin{equation}
  P_{qs}(\vec{r},\vec{r}')
=
  \langle \Psi|a^{\dagger}_{\vec{r}s q}a^{\dagger}_{\vec{r}'s q}
  a_{\vec{r}'s q}a_{\vec{r}s q}|\Psi \rangle.
\end{equation}
For the HF product state $|\Psi \rangle$  this probability can be written as
\begin{equation}
  P_{qs}(\vec{r},\vec{r}') =
  \rho_q(\vec{r}s, \vec{r}s)\rho_q(\vec{r}'s, \vec{r}'s)
  -
|\rho_q(\vec{r}s,\vec{r}'s)|^2.
\end{equation}
Because of the Pauli exclusion principle, $ P_{qs}(\vec{r},\vec{r})=0$.
If a nucleon with spin $s$ and isospin $q$
is located with certainty at position $\vec{r}$,  the conditional probability
of finding a second nucleon  with the same
spin and isospin at position $\vec{r}'$ is
\begin{equation}
\label{conditionalProb}
  R_{qs}(\vec{r},\vec{r}')
  =
 \frac{ P_{qs}(\vec{r},\vec{r}')}{\rho_q(\vec{r}s, \vec{r}s)}.
\end{equation}

To study the local (short-range) behavior of $R_{qs}$, one assumes that the second nucleon is located within a shell of small radius $\delta$  around $\vec{r}$. The corresponding conditional probability (\ref{conditionalProb}) can be written as
\begin{equation}
R_{qs}(\boldsymbol{r}, \boldsymbol{r}+\boldsymbol{\delta})=\left. e^{\boldsymbol{\delta}  \cdot\vec{\nabla}'} R_{qs}(\boldsymbol{r}, \boldsymbol{r}') \right|_{\boldsymbol{r}=\boldsymbol{r}'}.
\end{equation}
After performing an angular 
  averaging  over the $\delta$-shell and carrying out a Taylor expansion in $\delta$, one obtains:
\begin{equation}
\begin{aligned}
\left\langle e^{\vec{\delta} \cdot \vec{\nabla}'}\right\rangle &=\frac{1}{4 \pi} \int e^{\boldsymbol{\delta} \cdot \vec{\nabla}'} d \Omega \\
&=1+\frac{1}{3 !} \delta^{2} \nabla'^{2}+\frac{1}{5 !} \delta^{4} \nabla'^{4}+\cdots. 
\end{aligned}
\end{equation}
The resulting local probability becomes
\begin{equation}
R_{qs}(\boldsymbol{r}, \delta) = \frac{1}{6}\delta^2 \left.\nabla'^2
R_{qs}(\boldsymbol{r}, \boldsymbol{r}') \right|_{\boldsymbol{r}=\boldsymbol{r}'}
+\mathcal{O}(\delta^4).
\end{equation}
By introducing  a localization measure
 $D_{qs}(\vec{r})$ through the relationship
\begin{equation}
R_{q s}(\boldsymbol{r}, \delta)=\frac{1}{3} D_{q s}(\boldsymbol{r}) \delta^{2}+\mathcal{O}(\delta^{4}), 
\end{equation}
one can capture the short-range limit of the conditional like-spin pair probability.

For a rotationally-invariant and spin-unpolarized system, $D_{qs}(\vec{r})$ is independent of the choice of the spin-quantization axis. However, for the deformed and rotating nuclei considered in this study, one has to consider three different measures $D_{qs_\mu}(\vec{r})$ with $\mu=x,y,z$.

If one chooses  $\mu$ axis as the spin-quantization axis, one can define three spin-dependent local densities:
\begin{subequations}\label{eq:density_relation}
\begin{align}
\rho_{q s_\mu}(\vec{r}) &= \frac{1}{2}\rho_q(\vec{r}) + \frac{1}{2}\sigma_\mu {s}_{q\mu}(\vec{r}), \\
\tau_{q s_\mu}(\vec{r}) &= \frac{1}{2}\tau_q(\vec{r}) + \frac{1}{2}\sigma_\mu {T}_{q\mu}(\vec{r}), \\
\vec{j}_{q s_\mu}(\vec{r}) &= \frac{1}{2}\vec{j}_q(\vec{r}) + \frac{1}{2}\sigma_\mu \mathbb{J}_q(\vec{r})\cdot \vec{e}_\mu,
\end{align}
\end{subequations}
where $\sigma_\mu =2s_\mu=\pm1$ and   $\vec{e}_\mu$ is the unit vector in the direction of  the $\mu$ axis. 
After straightforward  algebraic manipulations based on the density-matrix expansion technique \cite{Negele1972,Dobaczewski2010}, the measure $D_{qs_\mu}(\vec{r})$ can be expressed through
the local densities (\ref{eq:density_relation}):
\begin{equation}\label{eq:prob_D}
D_{qs_\mu} = \tau_{qs_\mu}-\frac{1}{4}\frac{\left|\vec{\nabla}\rho_{qs_\mu}\right|^2}{\rho_{qs_\mu}}
-\frac{\left|\boldsymbol{j}_{qs_\mu}\right|^2}{\rho_{qs_\mu}}.
\end{equation}
Following Ref.~\cite{beckeSimpleMeasureElectron1990},
a dimensionless and normalized NLF can now be defined as
\begin{equation}\label{eq:NLF_def}
\mathcal{C}_{qs_\mu}(\boldsymbol{r})=\left[1+\left(\frac{D_{qs_\mu}(\vec{r})}{\tau_{qs_\mu}^{\mathrm{TF}}(\vec{r})}\right)^{2}\right]^{-1}, 
\end{equation}
where the normalization $\tau_{qs_\mu}^{\mathrm{TF}}(\vec{r})=\frac{3}{5}\left(6\pi^2\right)^{2/3}\rho_{qs_\mu}^{5/3}(\vec{r})$ is the Thomas-Fermi kinetic-energy density.


It should be noted that the densities (\ref{eq:density_relation}) constituting the NLF contain both time-even 
and time-odd components. Indeed, the particle density $\rho_q(\vec{r})$,  kinetic-energy density
$\tau_q(\vec{r})$, and  spin-current tensor density $\mathbb{J}_q(\vec{r})$ are all time-even, while the spin vector density ${s}_{q\mu}(\vec{r})$,  spin-kinetic vector density ${T}_{q\mu}(\vec{r})$, and current vector density $\vec{j}_q(\vec{r})$ are time-odd. 
If time-reversal symmetry is conserved,
${s}_{q\mu}(\vec{r})=0$,  ${T}_{q\mu}(\vec{r})=0$, and $\vec{j}_q(\vec{r})=0$.
Consequently, for a system that conserves  time-reversal symmetry and is governed by spin-independent interactions, one obtains:
\begin{equation}\label{Datrom}
 D_{q\pm\spin}= D_{q} =
\frac{1}{2}\tau_{q}-\frac{1}{8}\frac{\left|\vec{\nabla}\rho_{q}\right|^2}{\rho_{q}},
\end{equation}
which is the familiar atomic physics expression \cite{beckeSimpleMeasureElectron1990}.

In general, the  tensor density $\mathbb{J}_q(\vec{r})$ does not vanish even if the time-reversal symmetry is conserved~\cite{Rohozinski2010}. 
It can be decomposed into trace,
antisymmetric, and symmetric parts \cite{Perlinska2004}.
In many practical applications, the spin-current tensor is  approximated by its antisymmetric (spin-orbit current) part \cite{Vautherin1972}. However,  all components 
of $\mathbb{J}_q$ are important to characterize nuclear spin-orbit and tensor interactions \cite{Lesinski2007,Zalewski2008,Bender2009,Iwata2011} and the resulting spin polarization, which is
sensitive to spin saturation of nucleonic shells.
Consequently, the current $\vec{j}_{q s_\mu}(\vec{r})$ does not vanish even in  ground-state configurations of even-even nuclei. 
While its   contribution to the NLF was ignored in several previous calculations \cite{reinhardLocalizationLightNuclei2011, zhangNucleonLocalizationFragment2016,schuetrumpfClusterFormationPrecompound2017}, 
 the current-density contribution to the NLF practically vanishes in the nuclear interior, see discussion in Sec.~\ref{sec:results_density}. Consequently,
one can safely neglect this term when the goal is to use the NLF as a configuration-characterization tool.

The choice of the normalization function in Eq.~(\ref{eq:NLF_def}) is somehow arbitrary. In atomic physics applications and for time-reversal-invariant nuclear configurations, the density  $\rho_{qs_\mu}$ does not depend on spin.
In the general case, however,  nucleonic densities depend on the spin polarization. In this work, in order to emphasize the rotation-induced effects, we decided to stick to  the normalization function $\tau_{qs_\mu}^{\mathrm{TF}}$, which is different for spin-up and spin-down subsystems.

As discussed in Refs.\ \cite{savinElectronLocalizationSolidState1992, fuentealbaChapterUnderstandingUsing2007}, the localization function can also be interpreted in terms of the Pauli exclusion principle. Let us consider a situation, in which an isolated fermion of given spin $s$ and isospin $q$, is located in some region of space. The wave function of this particle can be written as
\begin{equation}
\psi_{qs}\left(\boldsymbol{r}\right)=\sqrt{\rho_{qs}}e^{i\chi\left(\boldsymbol{r}\right)},
\end{equation}
where $\chi\left(\boldsymbol{r}\right)$ is a position-dependent phase factor related to the current density via
\begin{equation}
\boldsymbol{j}_{qs} = \rho_{qs} \vec{\nabla} \chi. 
\end{equation}
The corresponding s.p.\ kinetic-energy density is the sum of last two terms in $D_{qs}$ (\ref{eq:prob_D}): 
\begin{equation}
\tau_{qs}^{\rm s.p.}=\left|\vec{\nabla} \psi_{qs}\right|^2 = \frac{1}{4}\frac{\left|\vec{\nabla} \rho_{qs}\right|^2}{\rho_{qs}} + \frac{\left|\boldsymbol{j}_{qs}\right|^2}{\rho_{qs}}, 
\end{equation}
where the first term is the von Weizsacker kinetic-energy density \cite{weizsaeckerZurTheorieKernmassen1935}.  
Therefore, $D_{qs}$ can be interpreted as   a measure of the excess
of kinetic-energy density due to the Pauli exclusion principle:
\begin{equation}\label{texcess}
D_{qs}=\tau_{qs}-\tau_{qs}^{\rm s.p.}.
\end{equation}
This interpretation of the NLF is more flexible as it does not involve the notion of the conditional probability (\ref{conditionalProb}), which is not straightforwardly generalized to the case of point-group symmetries of the nuclear mean field.

\subsection{Cranked Hartree-Fock calculations}\label{sec:crank}

Superdeformed nuclei around $^{152}$Dy can be viewed as unique laboratories of extreme single-particle behavior
\cite{satulaAdditivityQuadrupoleMoments1996,Clark2001}.
 The nucleus $^{152}$Dy  plays a role of superdeformed double-magic core due to large
shell closures at $Z=66$ and $N=86$. 
Because of this, $^{152}$Dy has been a subject of many studies of self-consistent nuclear response to collective rotation; see, e.g.,  Ref.~\cite{dobaczewskiTimeoddComponentsMean1995,Bonche1996,afanasjevTimeoddMeanFields2010,Hellemans2012}. 
Because of large deformed gaps  and rapid rotation,  pairing correlations are  weak in
SD $^{152}$Dy \cite{Shimizu1987,Baktash1995}. Indeed  with a reasonable pairing strength, adjusted to experimental odd-even mass difference in $^{120}$Sn as done in Ref.~\cite{Dobaczewski1995}, the static
pairing vanishes in the SD yrast band of $^{152}$Dy in Hartree-Fock-Bogoliubov (HFB) calculations.

The intrinsic configurations of SD bands in the $A=150$ mass region  are well characterized by nucleons in the intruder orbitals carrying large principal harmonic-oscillator (HO) numbers $\cal N$,
namely the proton ${\cal N}=6$ and neutron ${\cal N} =7$ states
\cite{Bengtsson1988,Nazarewicz1989}.
Because of their large intrinsic angular momenta, these orbitals strongly respond to nuclear rotation; hence, their occupations and alignment patterns well characterize SD bands.

To study the impact of rotation on shell structure through the nucleon localizations, we
carry out unpaired cranked HF (CHF) calculations for  superdeformed $^{152}$Dy
using  the HF solver {\sc hfodd} \cite{schunckSolutionSkyrmeHartree2012}.
Following  Ref.\ \cite{dobaczewskiTimeoddComponentsMean1995},
s.p.\ wave functions have been expanded in a stretched deformed HO basis 
with frequencies $\hbar\omega_z = 6.246$\,MeV  and $\hbar\omega_{\perp} = 11.200$\,MeV along the directions parallel and perpendicular to the symmetry axis, respectively. 
The total number of basis states is 1013 with HO quanta not exceeding 15 in each direction. We employed
the Skyrme energy density functional parametrization SkM$^*$ \cite{bartelBetterParametrisationSkyrmelike1982}, 
with its generic time-odd terms \cite{dobaczewskiTimeoddComponentsMean1995, benderGamowTellerStrengthSpinisospin2002}. 

The   angular momentum has been generated by means of a cranking term $-\omega \hat{J}_y$, where $\hat{J}_y$ is the $y$-component of the total  angular-momentum operator and $\omega$ represents the angular velocity of rotation. 
In the presence  of the cranking term, parity ($\hat{P}$), $y$ signature
($\hat{R}_y=\exp(-i\pi \hat{J}_y)$),   and $y$ simplex  ($\hat{\mathfrak R}_y=\hat{P}\hat{R}_y$) symmetries are preserved while  time-reversal and axial symmetries are broken; see
Refs.~\cite{Goodman1974,Dobaczewski1997,Dobaczewski2000} for more discussion. 
Since the time-reversal operator  commutes with the signature and simplex operators,
the time-reversed s.p.\ CHF states (Routhians) belong to opposite signature and simplex eigenvalues.

Every  CHF configuration can be labeled by using the standard notation in terms of parity-signature blocks $[N_{+,+i}, N_{+,-i}, N_{-,+i},N_{-,-i}]$, where
$N_{\pi r_y}$ are the numbers of occupied s.p.\ orbitals having 
parity $\pi$ and $y$ signature $r_y$.
As discussed in Ref.\ \cite{dobaczewskiTimeoddComponentsMean1995}, the yrast  configuration of SD $^{152}$Dy is  $[22,22,21,21]_n\otimes[16,16,17,17]_p$. 
The relative variation of the quadrupole moment $Q_{20}$ within this state is much less than 1\% in the frequency range $\hbar\omega=0.2 \sim 0.5$\,MeV \cite{satulaAdditivityQuadrupoleMoments1996}, so we constrain it at the value $Q_{20}=42$\,b to eliminate its possible impact on the computed localizations.

\begin{figure}[htb]
	\includegraphics[width=0.8\linewidth]{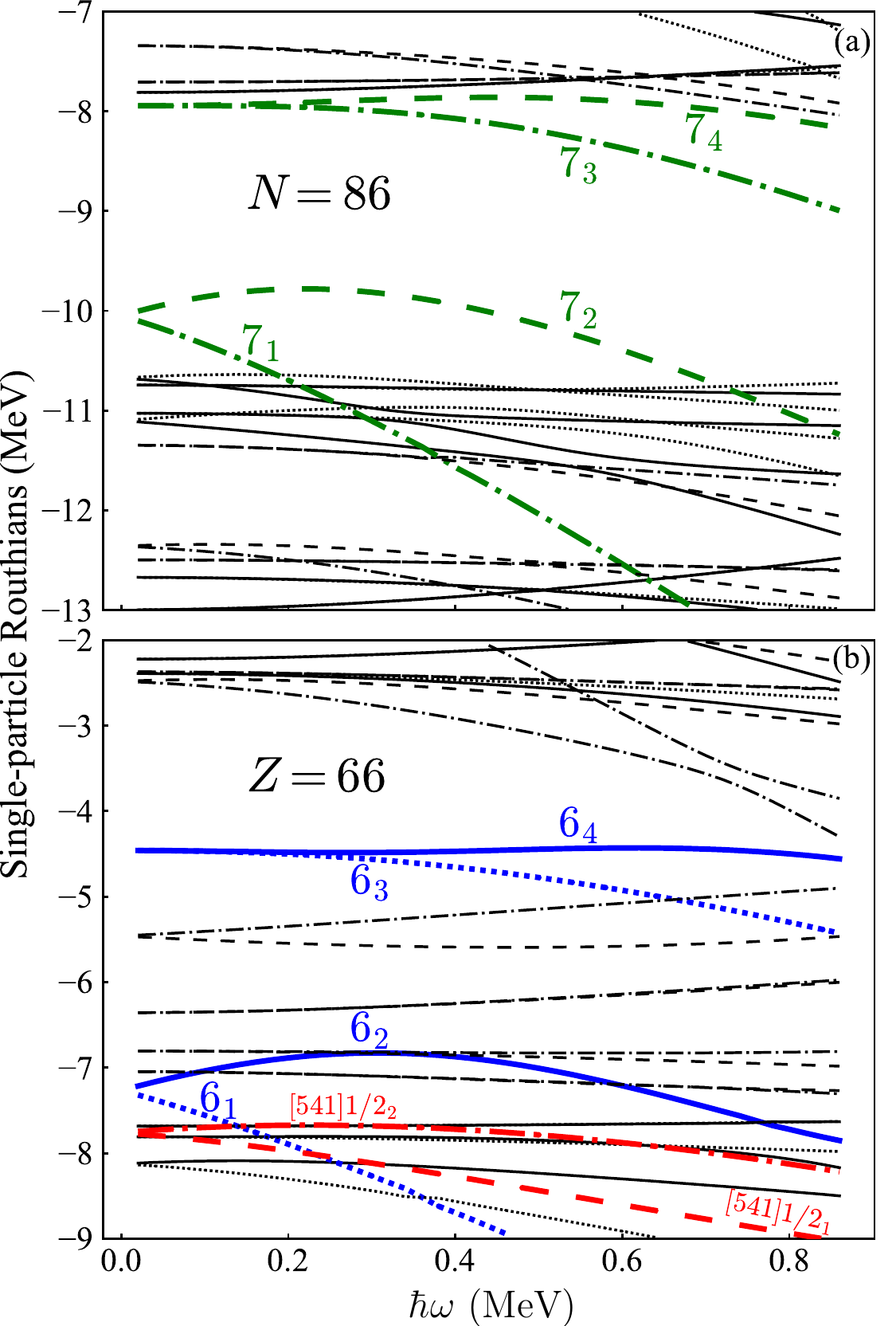}
	\caption{Single-particle (a) neutron and (b) proton Routhians as functions of $\omega$, obtained in the CHF+SkM$^{*}$ calculations for the SD yrast band of $^{152}$Dy. 
The ($\pi r_y$) combinations are indicated by solid lines ($+,+i$), dotted lines ($+,-i$),
dot-dashed lines ($-,+i$), and dashed lines ($-,-i$).  
The  Routhians originating from the lowest  neutron ${\cal N}=7$ and proton  ${\cal N}=6$
and [541]1/2  levels are marked by thicker lines.  
	}
	\label{fig:dy152_levels}
\end{figure}
Single-particle  Routhians obtained in the CHF+SkM$^{*}$ calculations for the SD yrast band of $^{152}$Dy are shown in Fig.\ \ref{fig:dy152_levels}. The large deformed shell closures   at $Z=66$ and $N=86$  are clearly seen.  The lowest
${\cal N}=7$ neutron and  ${\cal N}=6$ proton Routhians indicated in the figure are rotation aligned, i.e., they are strongly impacted by the Coriolis coupling and their s.p.\ aligned angular momenta are large at high rotational frequencies. Many  other states around the Fermi level  are weakly impacted by rotation. Such states are usually referred to as deformation-aligned (strongly coupled) \cite{Stephens1975,Boh75,Nil95aB}.

\subsection{Cranked harmonic-oscillator calculations}\label{sec:cho}

In the previous study of the NLF, the harmonic-oscillator model was used to provide an illustrative guidance \cite{Schuetrumpf2017a}. 
In this work, we study the NLF patterns of the SD
cranked harmonic-oscillator (CHO) model with frequencies  $\omega_{\perp}=\omega_x=\omega_y=2 \omega_z$. Since the HO potential is spin-independent,
every s.p.\ HO level is doubly-degenerate. As in the CHF calculations, we assume that the rotation takes place around the $y$ axis.
The s.p.\ Routhians and wave functions of the CHO can be obtained analytically
\cite{glasCrankedHarmonicOscillator1978,Troudet1981,szymanskiFastNuclearRotation1983}. 
We wish to emphasize that our CHO results were obtained without imposing
the consistency relation between mean-field ellipsoidal deformation and the average density distribution \cite{Boh75,Troudet1981}.

\begin{figure}[htb]
	\includegraphics[width=0.8\linewidth]{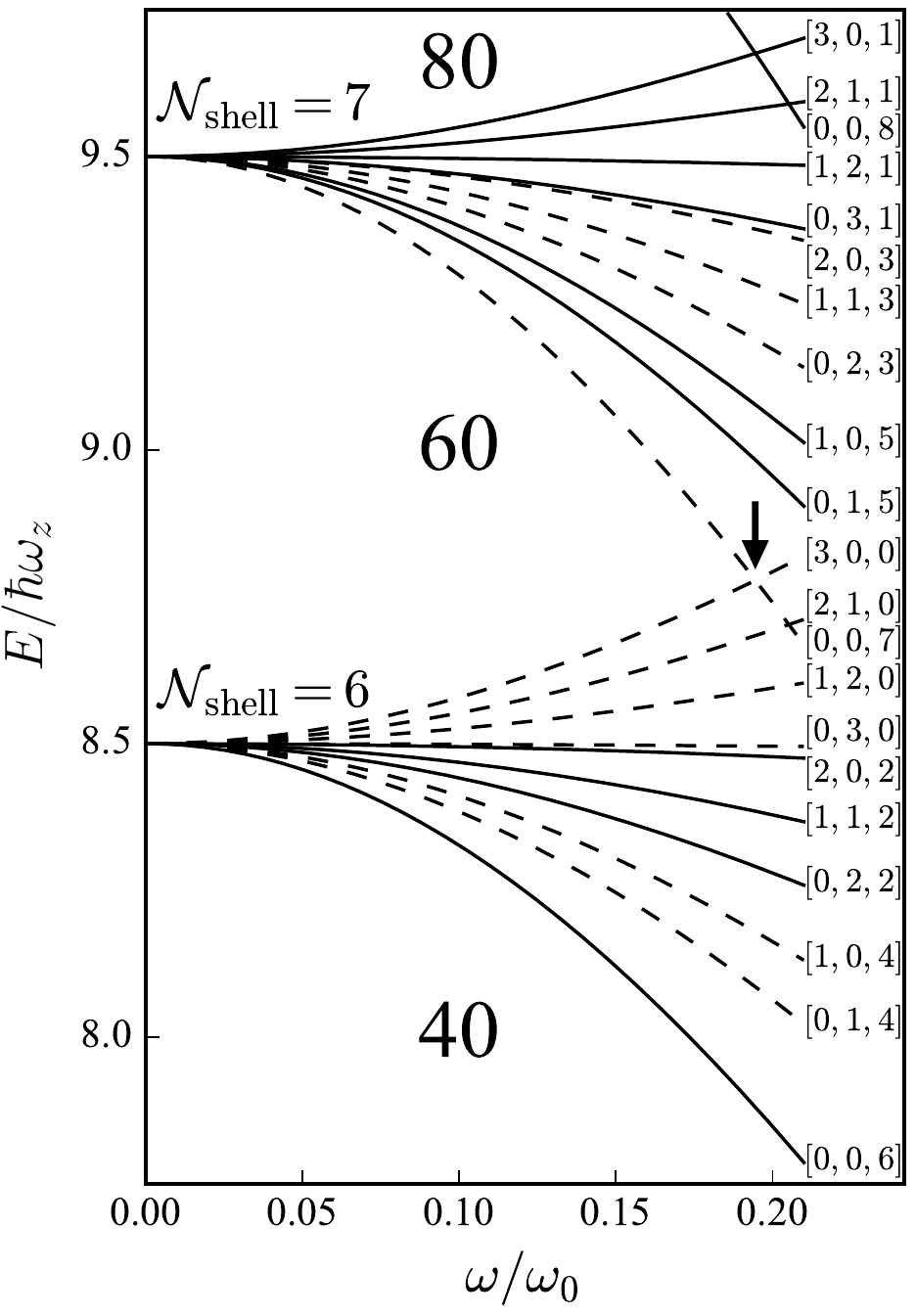}
	\caption{Single-particle Routhians of the SD CHO model belonging to the supershells
${\cal N}_{\rm shell}=6$ and 7.	
		The CHO quantum numbers $[n_1,n_2,n_3]$ are given in brackets.  Positive-parity and negative-parity states are indicated by solid  and dashed lines, respectively. 
		The rotational frequency $\omega$ is expressed in units of $\omega_0 = \left(\omega_z\omega^2_\perp\right)^{1/3}$ while  the Routhian $E$ is in units of $\hbar\omega_z$. Each level is doubly degenerate due  to the two possible spin orientations. The crossing between the lowest ${\cal N}=7$ Routhian [0,0,7] and the [3,0,0] Routhian at $\omega/\omega_0 \approx 0.2$ is marked by the arrow.
		}
	\label{fig:cho_levels}
\end{figure}

To relate the CHO analysis to the CHF results for  SD $^{152}$Dy, we study a  SD HO potential  filled with  60 particles, which corresponds to a closed SD supershell
${\cal N}_{\rm shell}\equiv 2(n_1+n_2)+n_3=6$
\cite{Boh75,Nazarewicz1992,Nil95aB,Szymanski1998}. The corresponding s.p.\ Routhians
are shown in Fig.~\ref{fig:cho_levels} as functions of $\omega$. 
A supershell of a SD HO consists of degenerate positive- and negative-parity states. 
This degeneracy is lifted by rotation: the orbits with no CHO quanta along the rotation axis ($n_2=0$) and the largest possible value of the difference $(n_3-n_1)$  carry the largest s.p.\ angular momentum. 
In Fig.~\ref{fig:cho_levels} those are the [0,0,7] (${\cal N}=7$)  and [0,0,6] (${\cal N}=6$)  Routhians.

\subsection{Nucleon localization function at high spins}\label{sec:NLFHS}

Since parity, $y$ signature $r_y$,  and $y$ simplex $\smy$ are self-consistent symmetries in our cranking calculations, in order to see the angular-momentum alignment effects caused by different orbits, it is convenient to study the  NLFs of a given  $r_y$ or  $\smy$.  This can be done by expressing local densities and currents in terms of their symmetry-conserving components. 
In practice, this can be done by summing up the contributions from HF s.p.\ wave functions belonging to a given symmetry block \cite{Goodman1974,Dobaczewski1997,Dobaczewski2000}.
For instance, if the $y$ simplex is conserved, 
\begin{equation}
\rho_{q}(\boldsymbol{r}) = \rho_{q \sms=+1}(\boldsymbol{r}) + \rho_{q \sms=-1}(\boldsymbol{r}),
\end{equation}
where $\breve{\sigma}_y \equiv  {\smy}/i=\pm 1$.
A  similar decomposition holds for $\tau_{q}(\boldsymbol{r})$ and $\boldsymbol{j}_{q}(\boldsymbol{r})$. 

By decomposing these densities into time-even and time-odd parts, they can be expressed in a form similar to Eq.\ (\ref{eq:density_relation}): 
\begin{subequations}\label{eq:density_relation_simplex}
	\begin{align}
	\rho_{q\sms}(\vec{r}) &= \frac{1}{2} \rho_q(\vec{r}) + \frac{1}{2} \breve{\sigma}_y s'_q(\vec{r}), \\
	\tau_{q\sms}(\vec{r}) &= \frac{1}{2} \tau_q(\vec{r}) + \frac{1}{2}\breve{\sigma}_y  T'_q(\vec{r}), \\
	\vec{j}_{q\sms}(\vec{r}) &= \frac{1}{2}\vec{j}_q(\vec{r}) +  \frac{1}{2} \breve{\sigma}_y \vec{J}_q'(\vec{r}),
	\end{align}
\end{subequations}
where 
\begin{subequations}
	\begin{align}
	s'_q(\vec{r}) &= \rho_{q\sms=+1}(\vec{r})-\rho_{q\sms=-1}(\vec{r}), \\
	T'_q(\vec{r}) &= \tau_{q\sms=+1}(\vec{r})-\tau_{q\sms=-1}(\vec{r}), \\
	\vec{J}'_q(\vec{r}) &= \vec{j}_{q\sms=+1}(\vec{r})-\vec{j}_{q\sms=-1}(\vec{r}). 
	\end{align}
\end{subequations}
The fields $s'$ and $T'$ are time-odd and $\vec{J}'$ is time-even.

\section{Results and discussion}\label{sec:results}

\subsection{General considerations}\label{sec:results_density}

\begin{figure}[htb]
	\includegraphics[width=\linewidth]{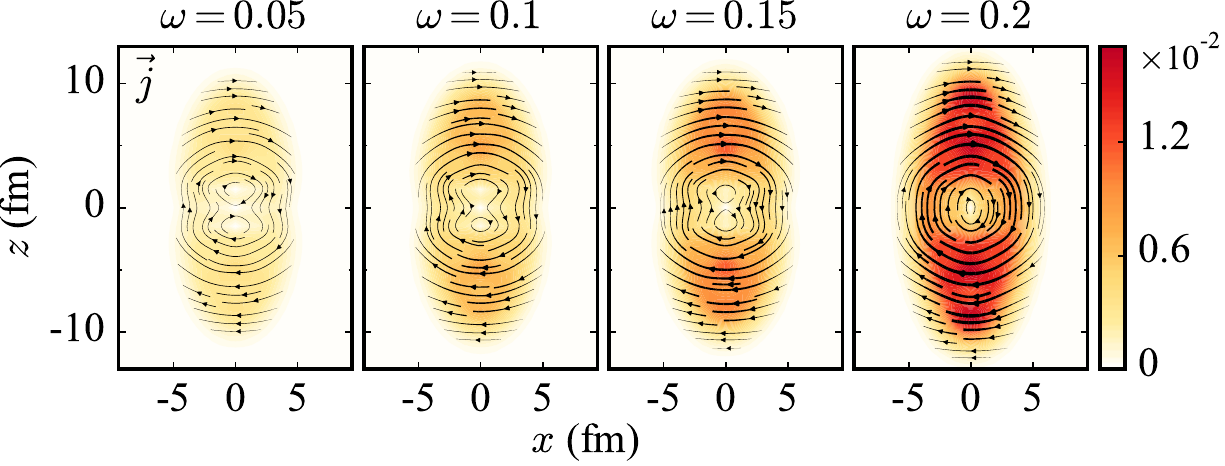}
	\caption{Current density $\boldsymbol{j}$ in the $x$-$z$ ($y=0$) plane, calculated in the CHO model with 60 particles in a SD HO well  for four values of rotational frequency $\omega$ (in units of $\omega_0$).   
		The magnitude $|\boldsymbol{j}|$ (in fm$^{-4}$) is shown by color and line thickness. }
	\label{fig:cho_current}
\end{figure}

In a rotating system, the  current density $\vec{j}$ characterizes the collective rotational behavior \cite{Radomski1976,Gulshani1978,Kunz1979,Fleckner1980,Durand1985,Laftchiev2003,Bartel2004,afanasjevTimeoddMeanFields2010,AfanasjevAbusara2018}.
Figure~\ref{fig:cho_current} shows how the current density  builds up in the CHO model. As rotational frequency increases,  a pattern of the vector field $\vec{j}$ resembling
 a rigid-body rotation gradually develops.
 At  $\omega=0.2\omega_0$, 
the lowest ${\cal N}=7$ Routhian [0,0,7] becomes occupied and  the [3,0,0] level becomes empty, see Fig.~\ref{fig:cho_levels}. 
As the orbital [0,0,7] is strongly prolate-driving and carries large s.p.\ angular momentum, and the Routhian [3,0,0] has large negative quadrupole moment (oblate), 
the associated  configuration change (band crossing) results in a large increase in the angular-momentum alignment and intrinsic deformation, see in  Fig.~\ref{fig:cho_current}. 
This effect is also present in CHO calculations which consider
the potential-density consistency relation \cite{Szymanski1998}.

\begin{figure}[htb]
	\includegraphics[width=\linewidth]{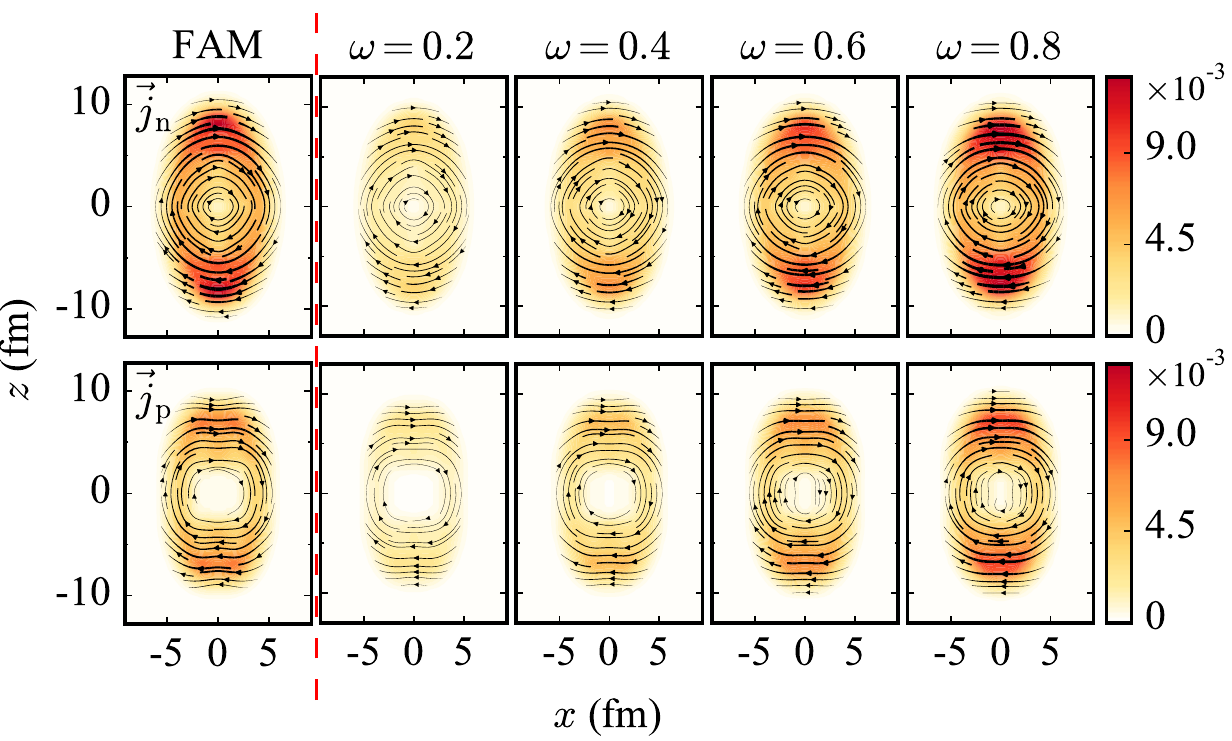}
	\caption{
		Current density $\boldsymbol{j}$ in the $x$-$z$ ($y=0$) plane for neutrons (top) and protons (bottom) in the SD yrast band of $^{152}$Dy obtained in the CHF  calculations, as a function of  $\omega$ (in units of MeV/$\hbar$). 
		The magnitude $|\boldsymbol{j}|$ (in fm$^{-4}$) is shown by color  and line thickness.
		The FAM-QRPA result is presented in the first column with a different color range. }
	\label{fig:dy152_current}
\end{figure}

When it comes to the realistic description, Fig.~\ref{fig:dy152_current} shows the neutron and proton current densities of $^{152}$Dy calculated in the CHF method at four rotational frequencies
up to  $\hbar \omega=0.8$\,MeV  (angular momentum $I_y\approx 90\,\hbar$).
The leftmost column in Fig.\ \ref{fig:dy152_current} shows the result of the benchmark quasiparticle random-phase approximation calculation using the finite amplitude method \cite{petrikThoulessValatinRotationalMoment2018}, which corresponds to the $\omega \rightarrow 0$   limit. 
Both FAM-QRPA and full cranking calculations produce flow patterns close to the rigid-body rotation. 
As irrotational flow originates from pairing correlations \cite{Fleckner1980,petrikThoulessValatinRotationalMoment2018} the result shown in Fig.~\ref{fig:dy152_current} is consistent with our assumption of no static pairing  in the SD yrast band of $^{152}$Dy.
 
\begin{figure}[htb]
	\includegraphics[width=\linewidth]{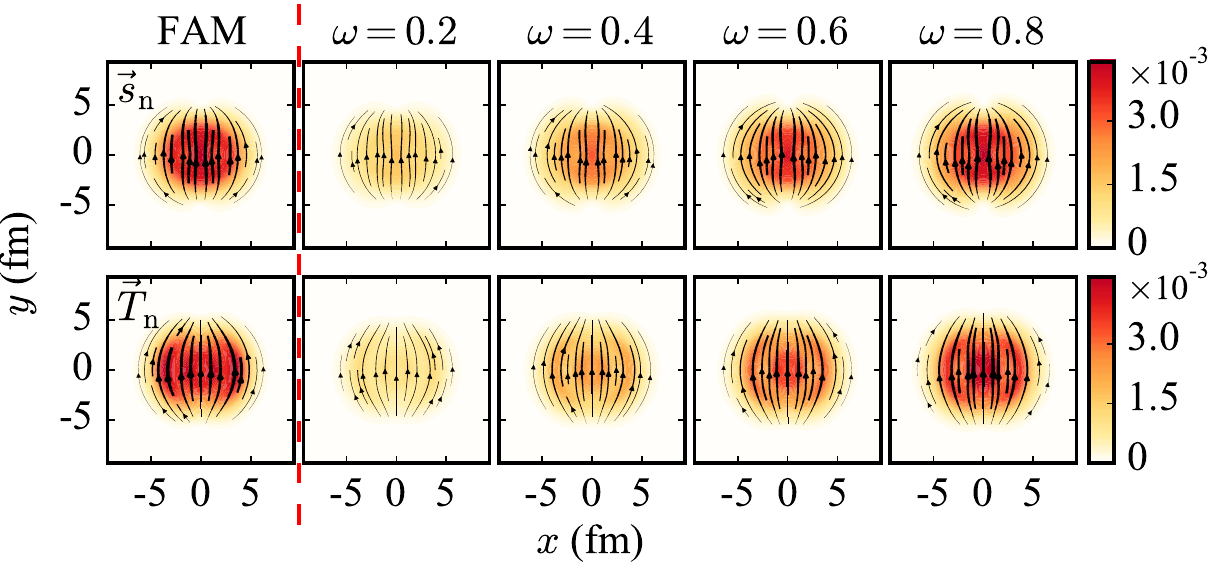}
	\caption{
		Spin density $\vec{s}$ (top) and spin-kinetic  density  $\vec{T}$ (bottom) in the $x$-$y$ ($z=0$) plane for neutrons in the SD yrast band of $^{152}$Dy obtained in the CHF  calculations, as functions of  $\omega$ (in units of MeV/$\hbar$). 
		The magnitudes, $|\vec{s}|$ (in fm$^{-3}$) and $|\vec{T}|$ (in fm$^{-5}$), are shown by color  and line thickness. The FAM-QRPA results are presented in the first column with a different color range. 
}
	\label{fig:dy152_spinfields}
\end{figure} 
In addition to the current $\vec{j}$, two other time-odd vector densities enter the expression for the NLF: spin density $\vec{s}$   and spin-kinetic  density  $\vec{T}$. They are displayed in Fig.~\ref{fig:dy152_spinfields} for several values of $\omega$. Both spin fields are polarized along the direction of the total angular momentum (here, $y$ axis). It is interesting to see that the distribution themselves hardly change with rotational frequency; what is changing is the magnitudes 
$|\vec{s}|$   and  $|\vec{T}|$ that gradually increase with rotation. This is also seen in the  FAM-QRPA calculation that  produces flow patterns close to those obtained in the CHF  calculations.

\begin{figure}[htb]
	\includegraphics[width=\linewidth]{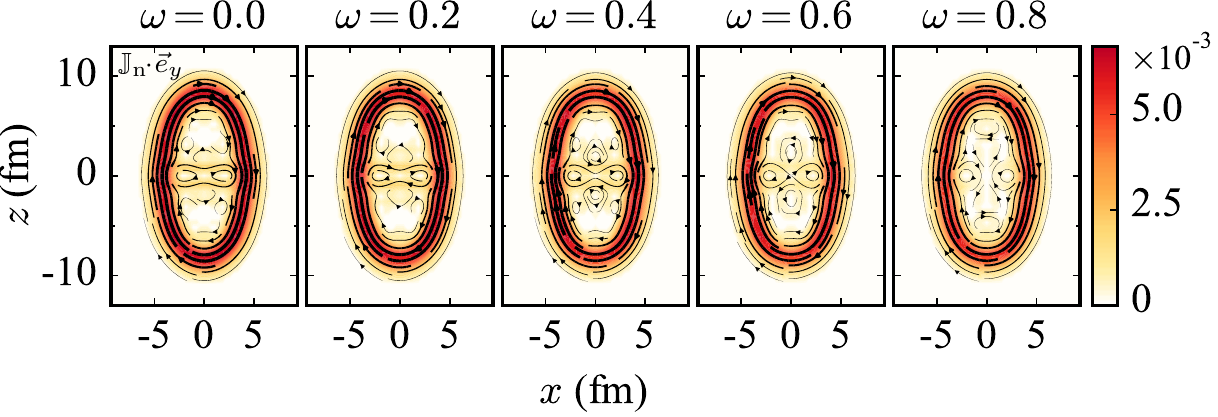}
	\caption{
		 Spin-current tensor density $\mathbb{J} \cdot \vec{e}_y$ in the $x$-$z$ ($y=0$) plane for neutrons in the SD yrast band of $^{152}$Dy, as a function of $\omega$ (in units of MeV/$\hbar$). 
		 Its magnitude (in fm$^{-4}$) is shown by color and line thickness.
}
	\label{fig:dy152_spintensor}
\end{figure}
To complete the discussion of spin fields, the  spin-current tensor density $\mathbb{J} \cdot \vec{e}_y$ is shown in Fig.~\ref{fig:dy152_spintensor}. As compared 
with the current density $\boldsymbol{j}$  shown in Fig.~\ref{fig:dy152_current}, $\mathbb{J} \cdot \vec{e}_y$ changes very weakly with $\omega$. 
This field has a surface character, i.e.,  it practically vanishes within the nuclear volume. Since $\mathbb{J}_q \cdot \vec{e}_y$ is time-even, its contribution to
$\mathcal{C}_{q s_\mu}$ does not vanish at $\omega=0$.

\subsection{Simplified nucleon localization function}\label{NLFt}

\begin{figure}[htb]
	\includegraphics[width=\linewidth]{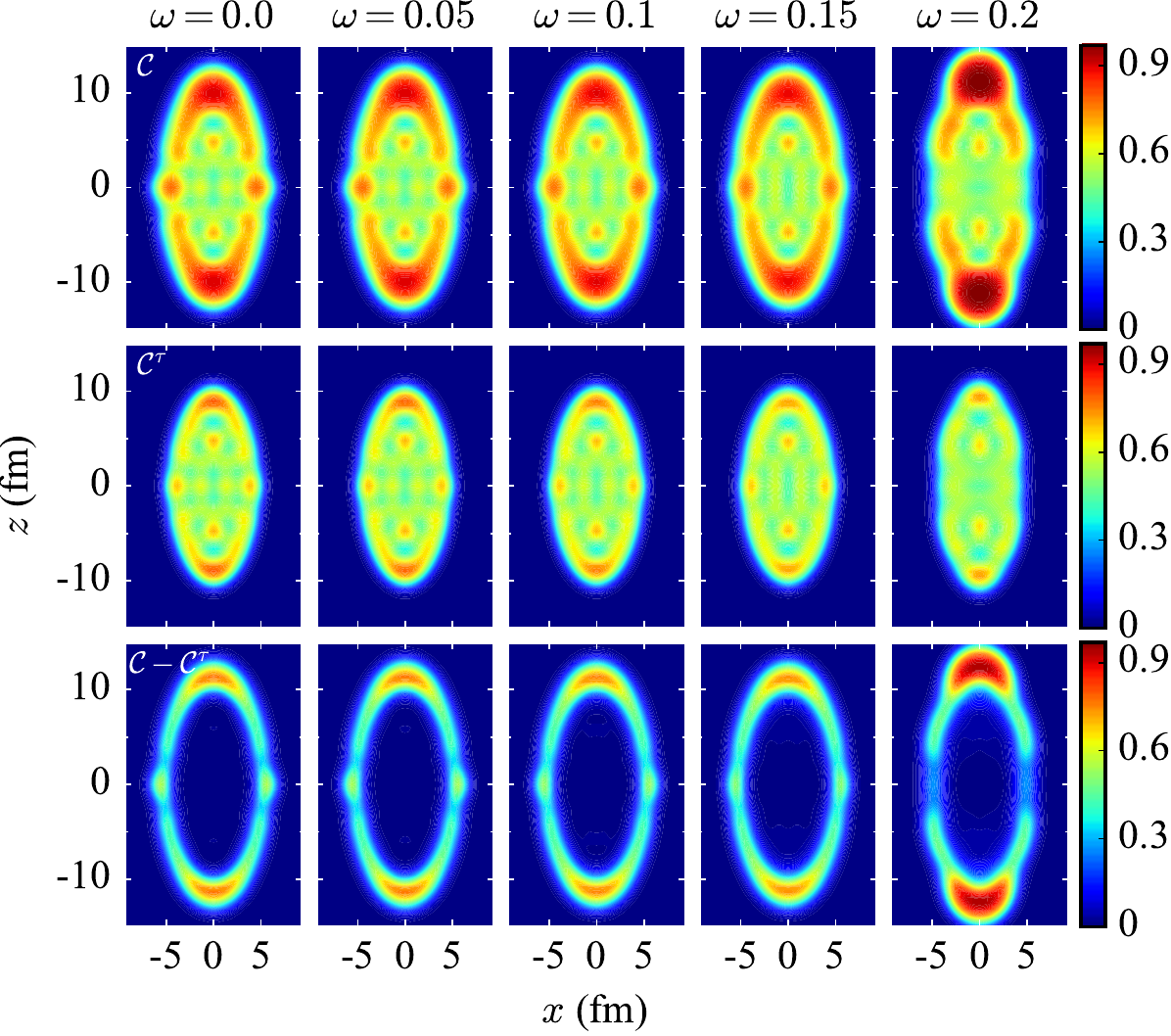}
	\caption{$\mathcal{C}$ (top), $\mathcal{C}^{\tau}$ (middle), and their difference (bottom) in the $x$-$z$ ($y=0$) plane, calculated in the CHO model with 60 particles in a SD HO well for five values of rotational frequency $\omega$ (in units of $\omega_0$).}
	\label{fig:cho_loc_comparison}
\end{figure}

\begin{figure}[htb]
	\includegraphics[width=\linewidth]{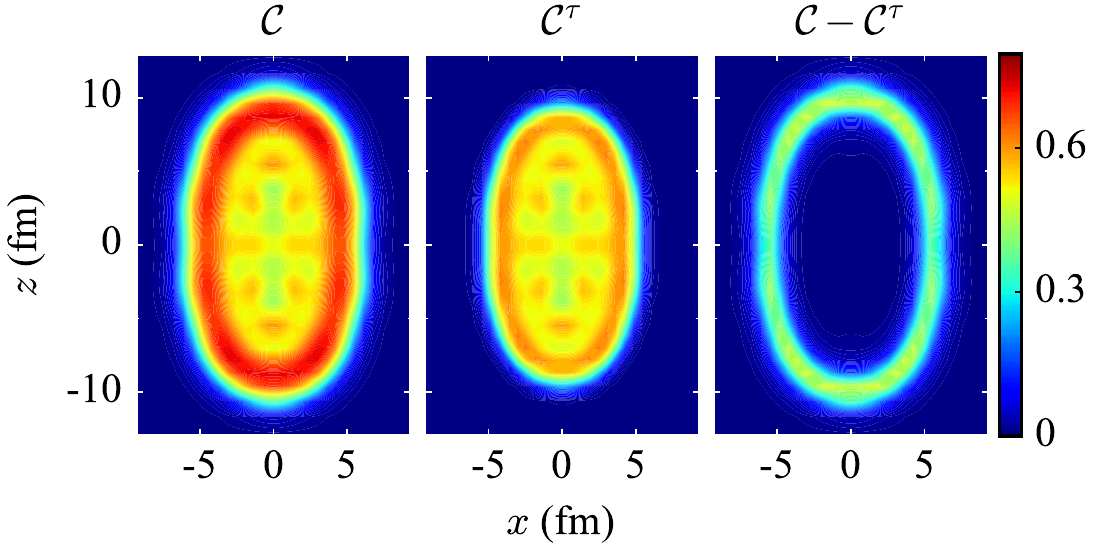}
	\caption{$\mathcal{C}$  (left), $\mathcal{C}^{\tau}$ (middle), and their difference (right)  in the $x$-$z$ ($y=0$) plane for neutrons with $\sms=-1$ ($y$ simplex $\smy=-1$), obtained in the CHF  calculations for the SD yrast configuration  of $^{152}$Dy at $\hbar\omega=0.9$\,MeV.
}
	\label{fig:dy152_loc_comparison}
\end{figure}

An important consequence of the rigid-body flow is that the current density only contributes significantly to the NLF at the surface. 
This observation should be valid in most cases even if an irrotational flow exists (see examples in Refs.\ \cite{Fleckner1980,petrikThoulessValatinRotationalMoment2018}). 
The same argument is also valid for the contribution to the NLF from the  density-gradient term $\left|\vec{\nabla} \rho_{q s}\right|^2$, which has a surface character.
Consequently, we define a simplified localization measure as
\begin{equation}\label{eq:NLFtau_def}
\mathcal{C}_{q s_\mu}^{\tau}(\vec{r})=\left[1+\left(\frac{\tau_{q s_\mu}(\vec{r})}{\tau_{q s_\mu}^{\mathrm{TF}}(\vec{r})}\right)^2\right]^{-1},
\end{equation}
which does not include contributions from the current density and density gradient. 
Figure~\ref{fig:cho_loc_comparison} shows $\mathcal{C}$, $\mathcal{C}^{\tau}$,  and their difference obtained in the CHO model; 
we indeed see that $\mathcal{C}^{\tau}$   exhibits the same pattern as  $\mathcal{C}$  inside the nuclear volume. A similar behavior is present in the CHF calculation for the SD yrast band of $^{152}$Dy.
Figure~\ref{fig:dy152_loc_comparison} shows  $\mathcal{C}$ and  $\mathcal{C}^{\tau}$ for neutrons with $\sms=-1$ ($y$ simplex $\smy=-i$)   at $\hbar\omega=0.9$\,MeV: 
the two localization functions   differ only in the surface region.
At lower frequencies, this difference is even less pronounced.

In previous work \cite{zhangNucleonLocalizationFragment2016} the NLF was normalized  as $\mathcal{C}_{q\sigma} \rightarrow \mathcal{C}_{q\sigma} \rho_{q\sigma}/\left[\max\rho_{q\sigma}\right]$ (with $\sigma$ being either spin, signature, or simplex) to avoid large values in the regions of small particle density. 
However, as shown in Figs.\ \ref{fig:cho_loc_comparison} and \ref{fig:dy152_loc_comparison}, replacing $\mathcal{C}$ with $\mathcal{C}^{\tau}$ mitigates this unwanted behavior and leaves the internal pattern unaffected, thus eliminating the need for this additional normalization.
Coming back to the interpretation of $D_{q s}$ as a measure of the Pauli
repulsion,
it is not surprising to see that $\left|\vec{\nabla}\rho_{q s}\right|^2$ and $\left|\boldsymbol{j}_{q s}\right|^2$ are significant only at the surface  
where only a limited number of s.p.\ orbits are available and thus become ``localized."
Therefore, the simplified localization function  $\mathcal{C}^{\tau}$ is a useful tool to characterize intrinsic configurations
in most cases, except  perhaps for  dynamic processes and high-energy modes  where the current density and density gradient can become appreciable  inside the nucleus.

\subsection{Dependence of nucleon localizations on the choice of spin quantization axis}\label{sec:smu}

\begin{figure}[htb]
	\includegraphics[width=\linewidth]{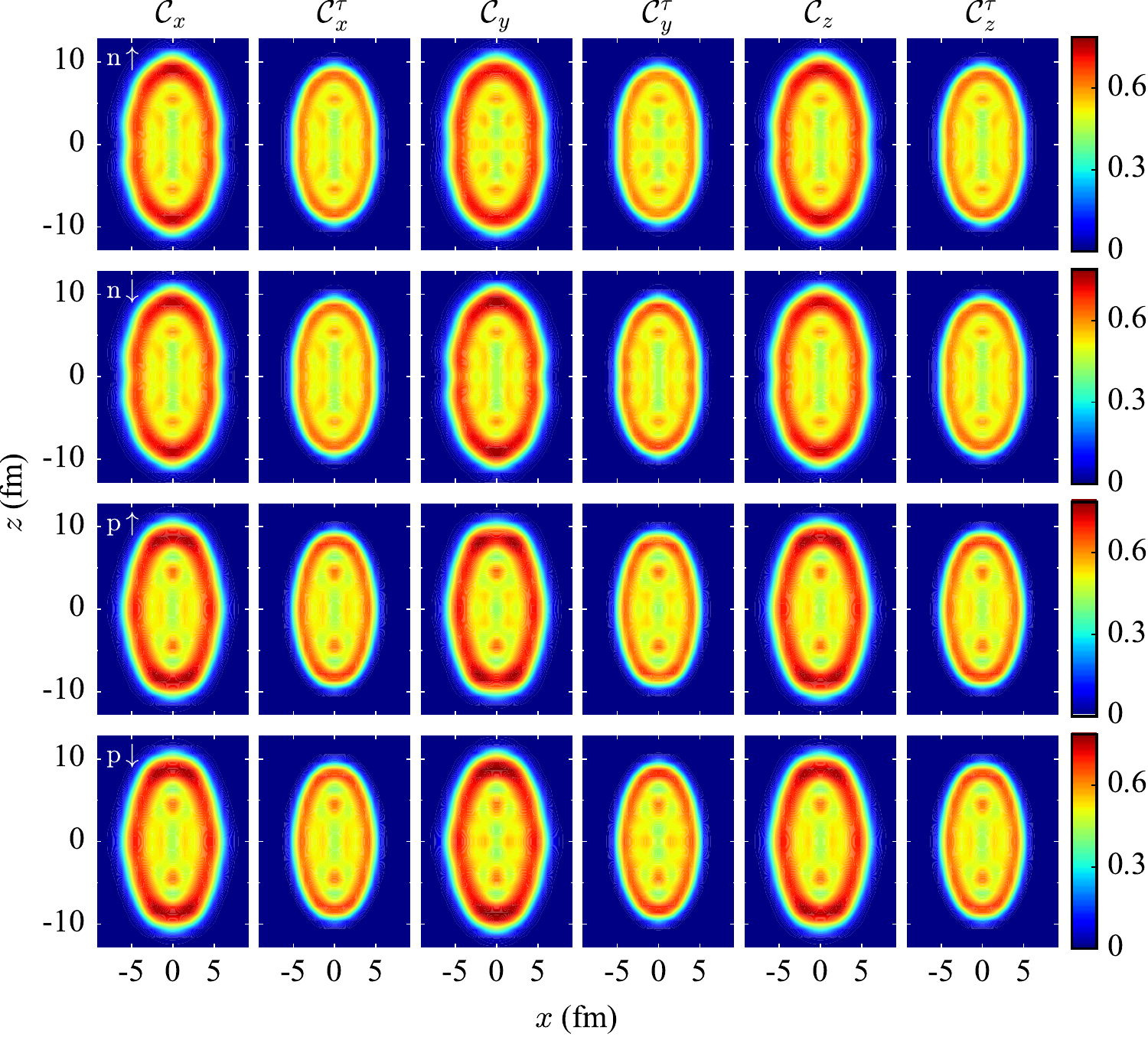}
	\caption{Nucleon localizations functions  $\mathcal{C}_{qs_\mu}$ (\ref{eq:NLF_def})
and  $\mathcal{C}^\tau_{qs_\mu}$ (\ref{eq:NLFtau_def})	
 in the $x$-$z$ ($y=0$) plane for three spin-quantization directions $\mu=x,y,z$, obtained in the CHF calculation for the SD yrast configuration  of $^{152}$Dy at $\hbar\omega=0.5$\,MeV.
}
	\label{fig:dy152_loc_xzplane}
\end{figure}

\begin{figure}[htb]
	\includegraphics[width=\linewidth]{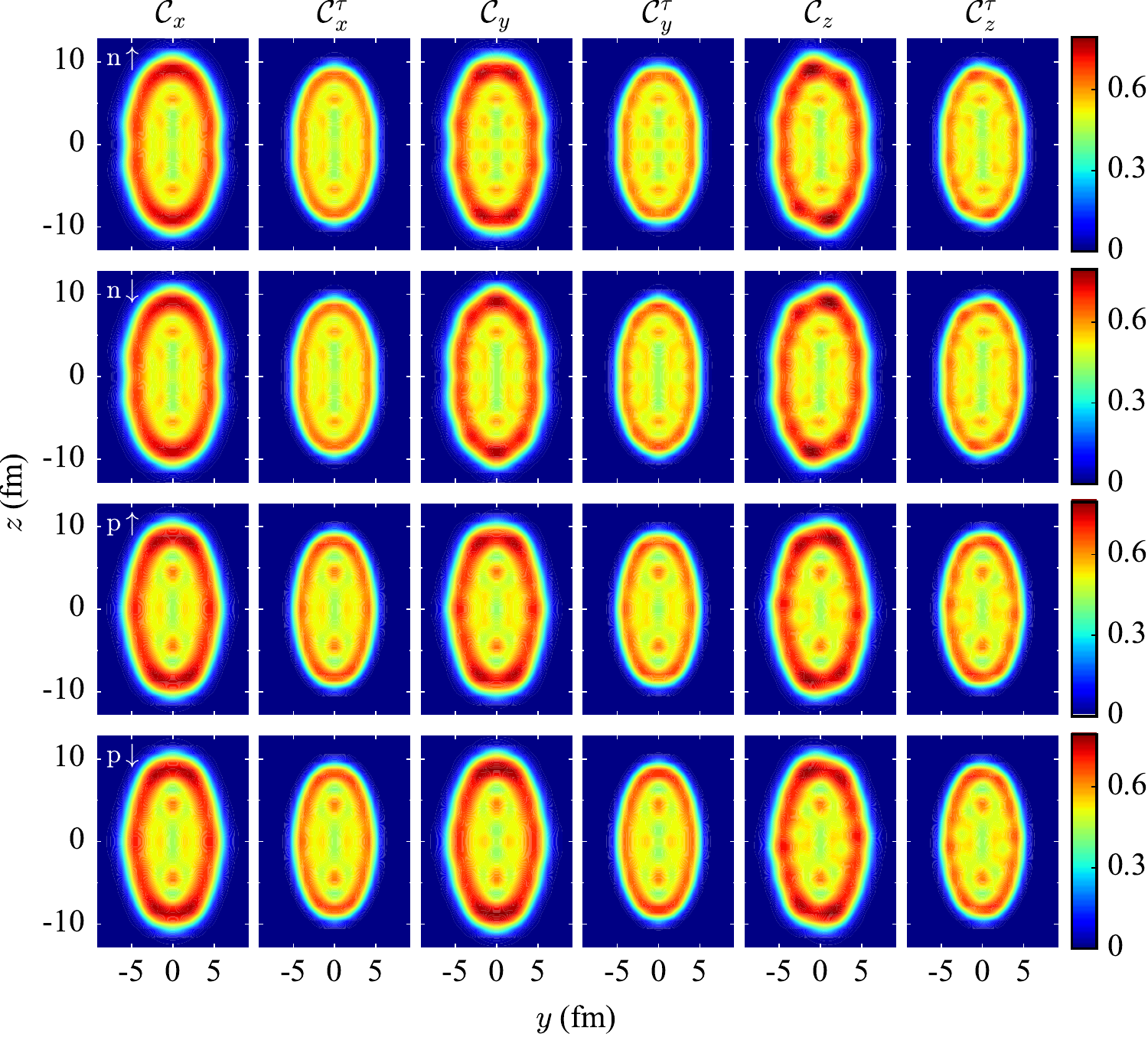}
	\caption{Similar to Fig.~\ref{fig:dy152_loc_xzplane} but shown in the
 $y$-$z$ ($x=0$) plane.
}
	\label{fig:dy152_loc_yzplane}
\end{figure}

As discussed in Sec.\ \ref{sec:NLF}, 
in the general case of deformed nuclei, nucleon localization functions  $\mathcal{C}_{qs_\mu}$ (\ref{eq:NLF_def}) and  $\mathcal{C}^\tau_{qs_\mu}$ (\ref{eq:NLFtau_def}) depend on the choice of  the spin-quantization direction $\mu$. This directional dependence is illustrated in Figs.~\ref{fig:dy152_loc_xzplane} and \ref{fig:dy152_loc_yzplane}
for the SD $^{152}$Dy at $\hbar\omega=0.5$\,MeV. 
It is seen that the NLF slightly depends on the choice of $\mu$, especially in the case of the $y$-$z$ cross section. More importantly,  $\mathcal{C}^\tau_{qs_\mu} \approx \mathcal{C}_{qs_\mu}$ in the nuclear interior, independently of $\mu$.

\subsection{Angular-momentum alignment: Cranked harmonic-oscillator analysis}\label{sec:results_ho}

In this section, we use the CHO model to illustrate some general features of NLFs and densities, which will help us understand the CHF results.
First, to show the usefulness of $\mathcal{C}^{\tau}$ when it comes to the visualization 
of nucleonic shell structure and angular-momentum alignment, we come back to
Fig.\ \ref{fig:cho_loc_comparison}. A characteristic regular pattern
 seen at $\omega=0$ gradually gets blurred with $\omega$.
At $\omega=0.2\omega_0$, where  the band crossing occurs,  $\mathcal{C}^{\tau}$ rapidly changes. Namely,
the number of maxima along the $z$ axis increases  as the [0,0,7] orbit becomes occupies, and the number of maxima along the $x$ axis decreases as the [3,0,0] state gets emptied.

\begin{figure}[htb]
	\includegraphics[width=\linewidth]{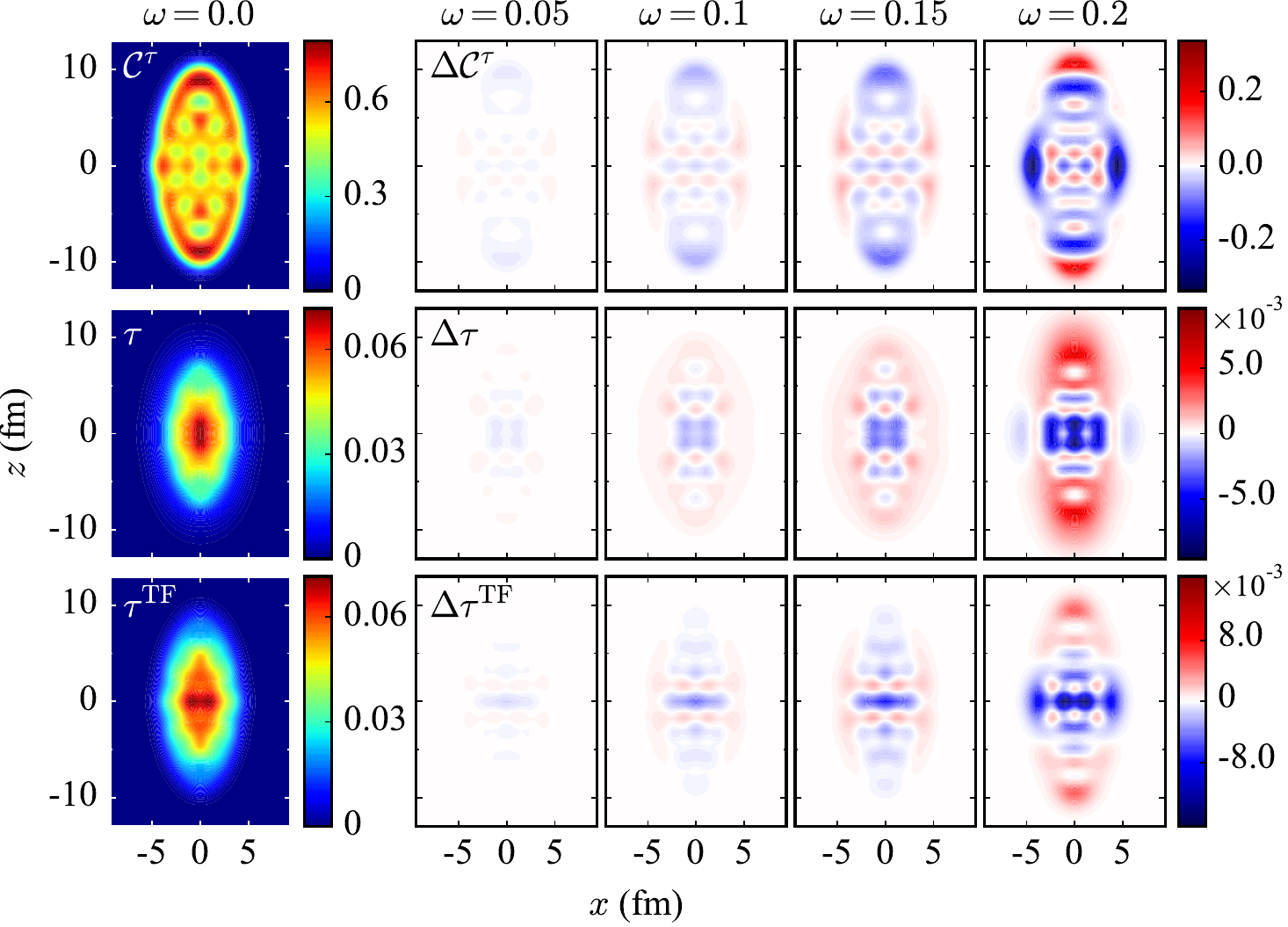}
	\caption{$\mathcal{C}^{\tau}$ (top),  $\tau$ (in fm$^{-5}$, middle) and  $\tau^{\mathrm{TF}}$  (in fm$^{-5}$, bottom) in the $x$-$z$ ($y=0$) plane, calculated in the CHO model with 60 particles in a SD HO well. 
		The first column shows  the reference plots  at $\omega=0$  while the other columns show the rotational dependence relative to the $\omega=0$ reference as a function of $\omega$ (in units of $\omega_0$). 
	}
	\label{fig:cho_NLFtau_tau_tauTF_diff_xz}
\end{figure}

To clearly see the evolution of $\mathcal{C}^{\tau}$  with $\omega$, we 
consider the indicator
\begin{equation}\label{relC}
\Delta \mathcal{C}^\tau(\vec{r};\omega)  \equiv \mathcal{C}^\tau (\vec{r};\omega) - \mathcal{C}^\tau (\vec{r};\omega=0).
\end{equation}
This quantity is shown in Fig.\ \ref{fig:cho_NLFtau_tau_tauTF_diff_xz} together with
the corresponding  variations  $\Delta\tau$ and  $\Delta\tau^{\mathrm{TF}}$ relative to the nonrotating case.

One can notice that there is a clear correspondence between the peaks of $\Delta \mathcal{C}^\tau$ and valleys (peaks) of $\Delta \tau$ ($\Delta \tau^{\mathrm{TF}}$), which is consistent with Eq.\ (\ref{eq:NLFtau_def}). 
This observation suggests that $\Delta \tau$ and $\Delta \tau^{\mathrm{TF}}$ are in antiphase, which results in a constructive interference when considering  their ratio.

\begin{figure}[htb]
	\includegraphics[width=0.8\linewidth]{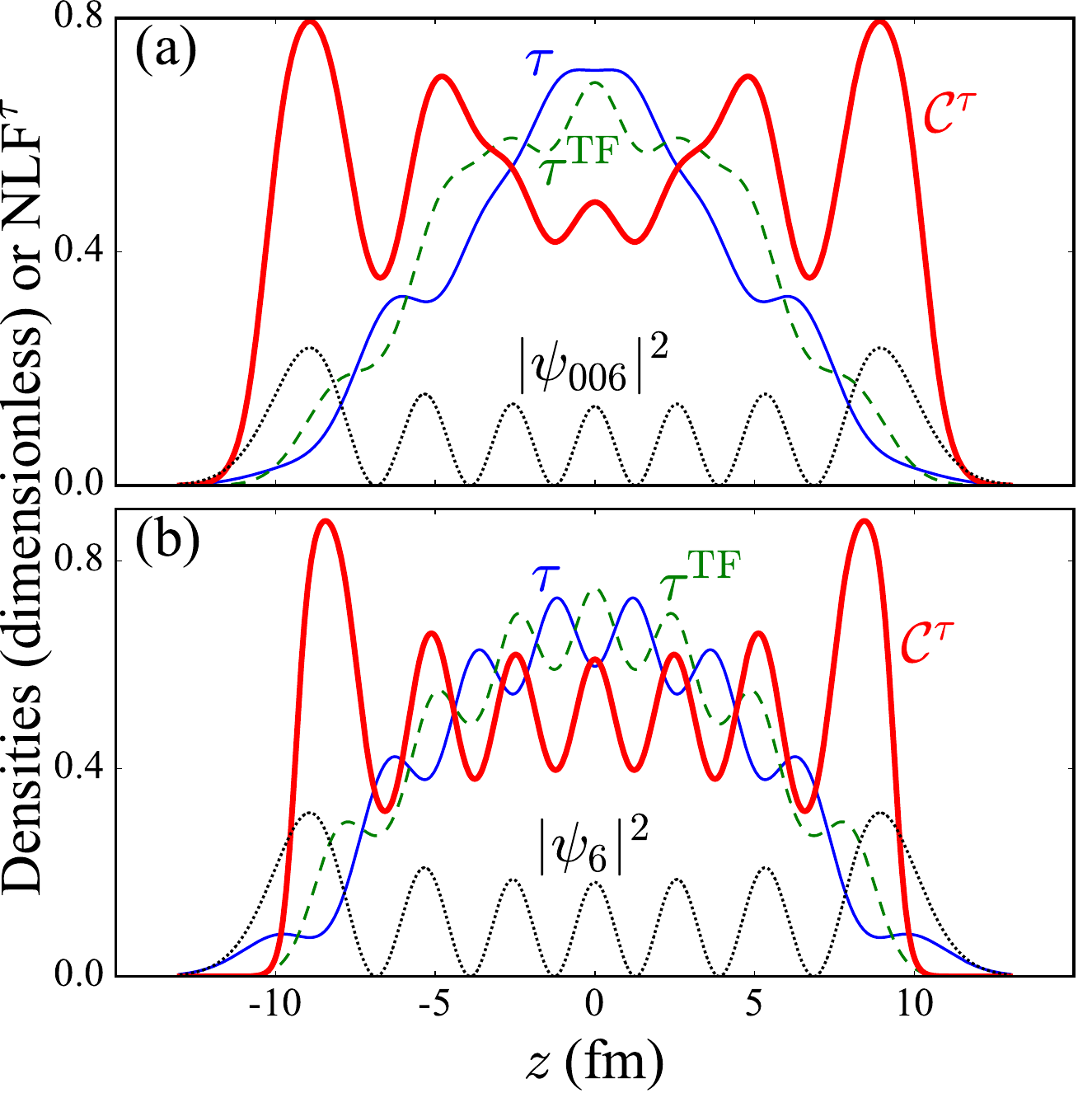}
	\caption{$\mathcal{C}^\tau$ (thick solid line), $\tau$ (solid line), and $\tau^{\mathrm{TF}}$ (dashed line)  for the nonrotating HO model plotted along $z$ axis ($x=y=0$). 
(a) Three-dimensional SD HO case with 60 particles.  The density profile of the [0,0,6] orbit is marked by a dotted line.   
(b) One-dimensional case. HO orbits with principal quantum number ${\cal N} \le 6$ are occupied. The density profile of the ${\cal N}=6$ orbit is marked by a dotted line;  
 here $\tau^{\mathrm{TF}}=\pi^2\rho^3/3$.  Some quantities are scaled for a better visualization. }
	\label{fig:ho_NLF_tau_tauTF_sp}
\end{figure}

To analyze this pattern in more detail,
Fig.\ \ref{fig:ho_NLF_tau_tauTF_sp}(a) displays $\tau$, $\tau^{\mathrm{TF}}$, 
and $\mathcal{C}^\tau$ for 60 particles in the nonrotating SD HO  along the $z$ axis ($x=y=0$), together with the density profile of  the  [0,0,6] state.
One can see that  valleys (peaks) of $\tau$, $\tau^{\mathrm{TF}}$, and   $\mathcal{C}^\tau$ roughly coincide with  maxima of the  [0,0,6] density, while other  states contribute to a smooth background. 

\begin{figure*}[htb]
	\includegraphics[width=\textwidth]{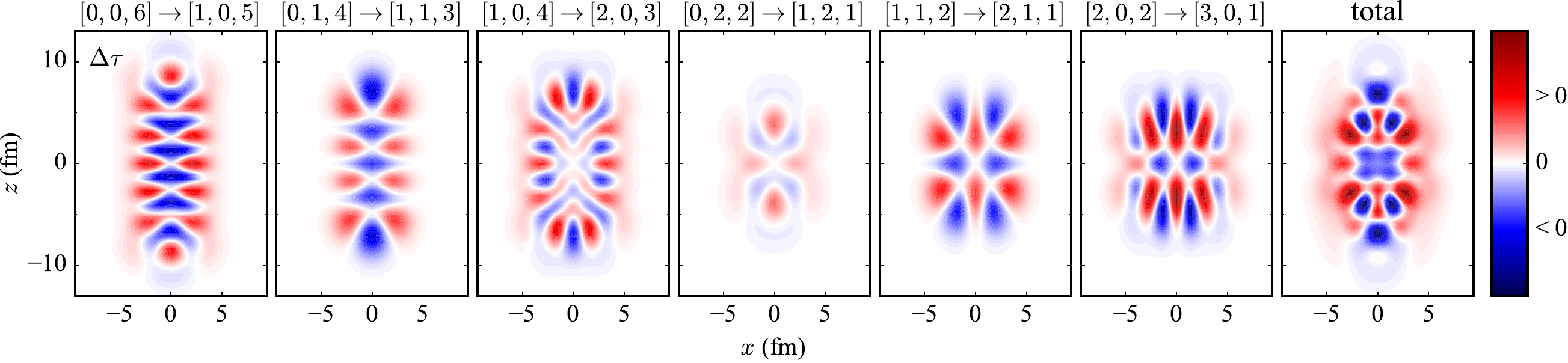}
	\caption{Changes in the kinetic-energy density $\tau$ due to   p-h excitations (at $\omega=0$)  from the  SD shell ${\cal N}_{\rm shell}=6$ to the next supershell
${\cal N}_{\rm shell}=7$ in Fig.~\ref{fig:cho_levels}. 
These excitations are induced in the CHO description of a  60-particle system  by the cranking term. The rightmost panel shows  the uniform average  of individual p-h contributions. }
	\label{fig:cho_sp_contribution_xz}
\end{figure*}

This effect is even more pronounced in the one-dimensional HO model, as shown in Fig.\ \ref{fig:ho_NLF_tau_tauTF_sp}(b) where HO orbits with quantum number ${\cal N} \le 6$ are occupied. 
The antiphase  relationship between  $\tau$ and $\tau^{\mathrm{TF}}$ is expected since $\tau$ is related to the gradients of s.p.\ wave functions while  $\tau^{\mathrm{TF}}$  depends on s.p.\ wave functions alone.
The advantage of $\mathcal{C}^\tau$  is that it amplifies the characteristic nodal structure of aligned  high-${\cal N}$ s.p.\ orbitals
thanks to the  constructive interference  between $\tau$ and $\tau^\mathrm{TF}$.

As discussed above, the kinetic-energy density $\tau$ is sensitive to the nodal structure of s.p.\ wave functions. This sensitivity can thus 
 be utilized for the visualization of the alignment process seen in  the pattern of $\Delta \tau$  in Fig.\ \ref{fig:cho_NLFtau_tau_tauTF_diff_xz}.  
(For discussion of quasimolecular states in light nuclei based on the  nodal structure of the s.p.\ densities and currents, see Ref.~\cite{AfanasjevAbusara2018}.)
The cranking operator 
$\omega \hat{L}_y$ induces the particle-hole (p-h) excitations
across the Fermi level.
The  low-energy  excitations  correspond to $\Delta{\cal N}=0$ ($\Delta n_1= \pm 1,\ \Delta n_2=0,\ \Delta n_3=\mp 1$) transitions.

Figure~\ref{fig:cho_sp_contribution_xz} shows the variation of $\tau$  at $\omega=0$ induced by six such p-h excitations across the $N=60$ gap 
from the  occupied supershell ${\cal N}_{\rm shell}=6$ to the empty supershell
${\cal N}_{\rm shell}=7$; see  Fig.~\ref{fig:cho_levels}. 
The [0,0,6]$\rightarrow$[1,0,5] excitation can be associated with that between the [660]1/2
($6_{1,2}$) and [651]3/2 ($6_{3,4}$) Nilsson levels. 
Both are rotation-aligned, prolate-driving orbits, and the corresponding $\Delta\tau$ plot exhibits a nodal pattern along the symmetry axis. 
On the other extreme, the  [2,0,2]$\rightarrow$[3,0,1] excitation corresponds to a [420]1/2 ([422]3/2)$\rightarrow$[411]3/2 ([413]5/2) transition, which involves  deformation-aligned orbits. 
The related $\Delta\tau$ plot exhibits a nodal pattern along the minor axis. 
By  summing up all six contributions, one arrives at a pattern in the last panel of Fig.~\ref{fig:cho_sp_contribution_xz}, which is indicative of a change in $\tau$ due to rotation. 
Interestingly, this pattern is quite similar to that of Fig.\ \ref{fig:cho_NLFtau_tau_tauTF_diff_xz} at $\omega=0.15\omega_0$.
We can thus conclude that, for a system that is strongly elongated  along $z$ axis, 
rotation-aligned s.p.\ states with large $n_3$ leave a strong imprint on $\Delta \tau$ and $\Delta \mathcal{C}^\tau$.

\subsection{Angular-momentum alignment: Cranked Hartree-Fock analysis}\label{sec:results_Dy152NLF}

\begin{figure}[htb]
	\includegraphics[width=\linewidth]{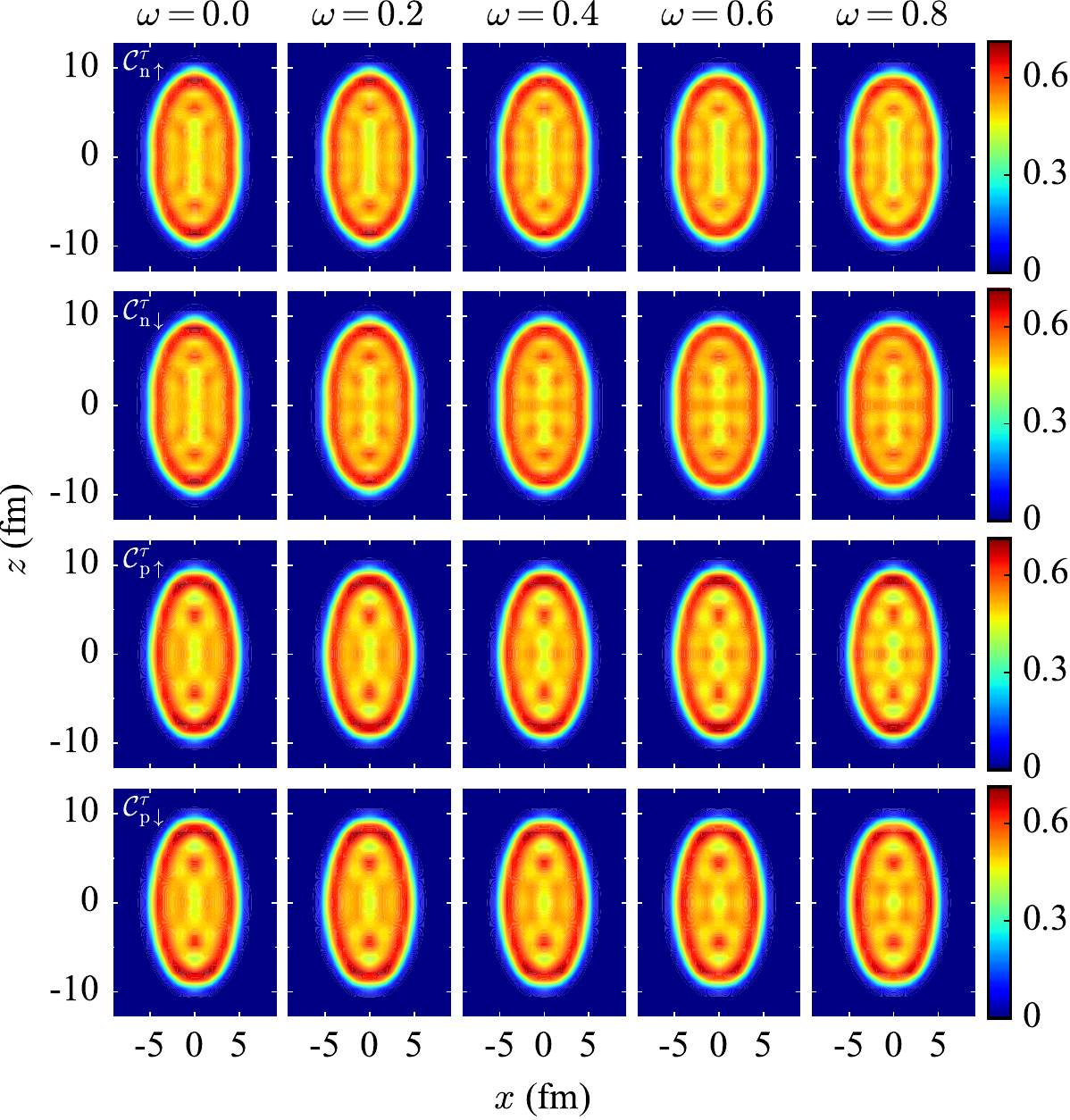}
	\caption{${\cal C}_{q\sms}^\tau$  in the $x$-$z$ ($y=0$) plane as a function of $\omega$ (in units of MeV/$\hbar$), obtained in the CHF calculation for the SD yrast band of $^{152}$Dy. The symbols $\uparrow$ and $\downarrow$ represent $\sms=+1$ and $-1$ ($y$ simplex $\smy=+i$ and $-i$), respectively.}
	\label{fig:dy152_NLFtau_xz}
\end{figure}

In this section, we study the localization patterns obtained in  the CHF  calculations  for the SD yrast band in   $^{152}$Dy. 
Figure~\ref{fig:dy152_NLFtau_xz} shows the simplified NLF ${\cal C}_{q\sms}^\tau$ in the $y=0$ plane for different 
values of $\omega$.
The first column corresponds to the nonrotating case, where we see  NLF patterns characteristic of a deformed nucleus, 
similar to those for $^{100}$Zr, $^{232}$Th, and $^{240}$Pu discussed in Ref.\ \cite{zhangNucleonLocalizationFragment2016}.
As $\omega$ increases, new patterns gradually emerge inside the nucleus, with $\mathcal{C}^\tau_{q\uparrow} \ne \mathcal{C}^\tau_{q\downarrow}$ due  to the time-reversal symmetry-breaking terms in (\ref{eq:density_relation_simplex}).

\begin{figure}[htb]
	\includegraphics[width=0.9\linewidth]{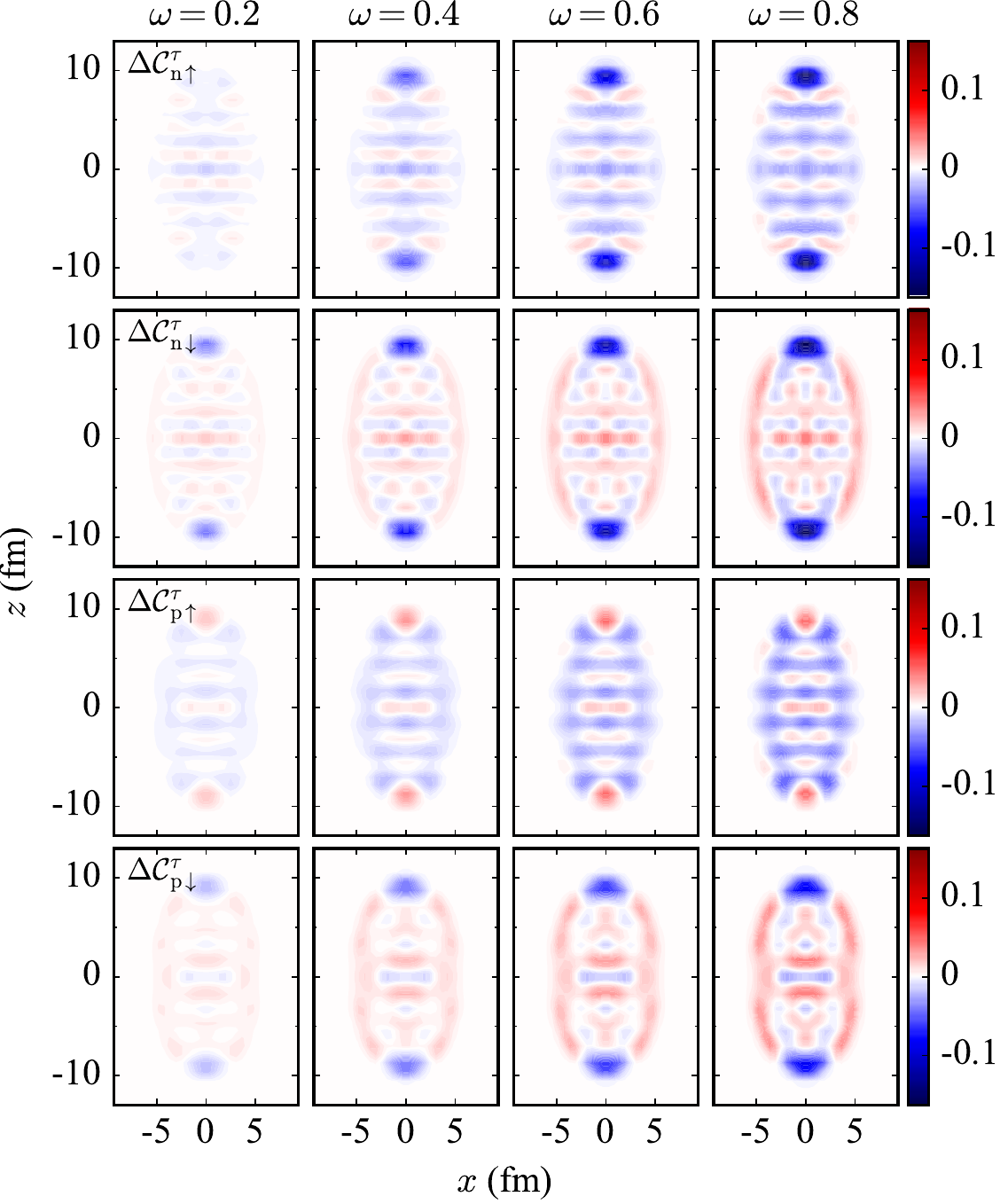}
	\caption{Similar to Fig.~\ref{fig:dy152_NLFtau_xz} but for $\Delta{\cal C}_{q\sms}^\tau$. The reference value of ${\cal C}^\tau$ at $\omega=0$ is shown in the first column of Fig.~\ref{fig:dy152_NLFtau_xz}.}
	\label{fig:dy152_NLFtau_diff_xz}
\end{figure}
For a better visualization of rotational dependence, we will be using relative  indicators, cf.\ Eq.~(\ref{relC}). 
Figure \ref{fig:dy152_NLFtau_diff_xz} presents the relative indicator $\Delta{\cal C}_{q\sms}^\tau$ in the  $y=0$  plane. 
As the local densities and currents can be decomposed into time-even and time-odd parts,
see Eqs.\ (\ref{eq:density_relation}) and (\ref{eq:density_relation_simplex}), 
their relative indicators can also be decomposed into time-even and time-odd components. For instance, 
\begin{equation}\label{eq:relative-density_relation_simplex}
\Delta\tau_{q\sms}(\vec{r};\omega) = \frac{1}{2} \Delta\tau_q(\vec{r};\omega) + \frac{1}{2}\sms  T'_q(\vec{r};\omega), 
\end{equation}
where the quantity 
\begin{equation}
\Delta\tau_q(\vec{r};\omega)=\tau_q(\vec{r};\omega)-\tau_q(\vec{r};\omega=0)
\end{equation}
does not depend on simplex and provides a background that is an even function of $\omega$. The 
simplex-dependent term in Eq.~(\ref{eq:relative-density_relation_simplex}) is $\omega$-odd; together with the time-odd component of
$\rho_{q\sms}(\vec{r};\omega)$ is responsible for the difference between the values of ${\cal C}_{q\sms}^\tau$ of  different simplex
(the same argument also holds for  signature). 
This difference is clearly shown in Fig.\ \ref{fig:dy152_NLFtau_diff_xz}. 

\begin{figure}[htb]
	\includegraphics[width=0.9\linewidth]{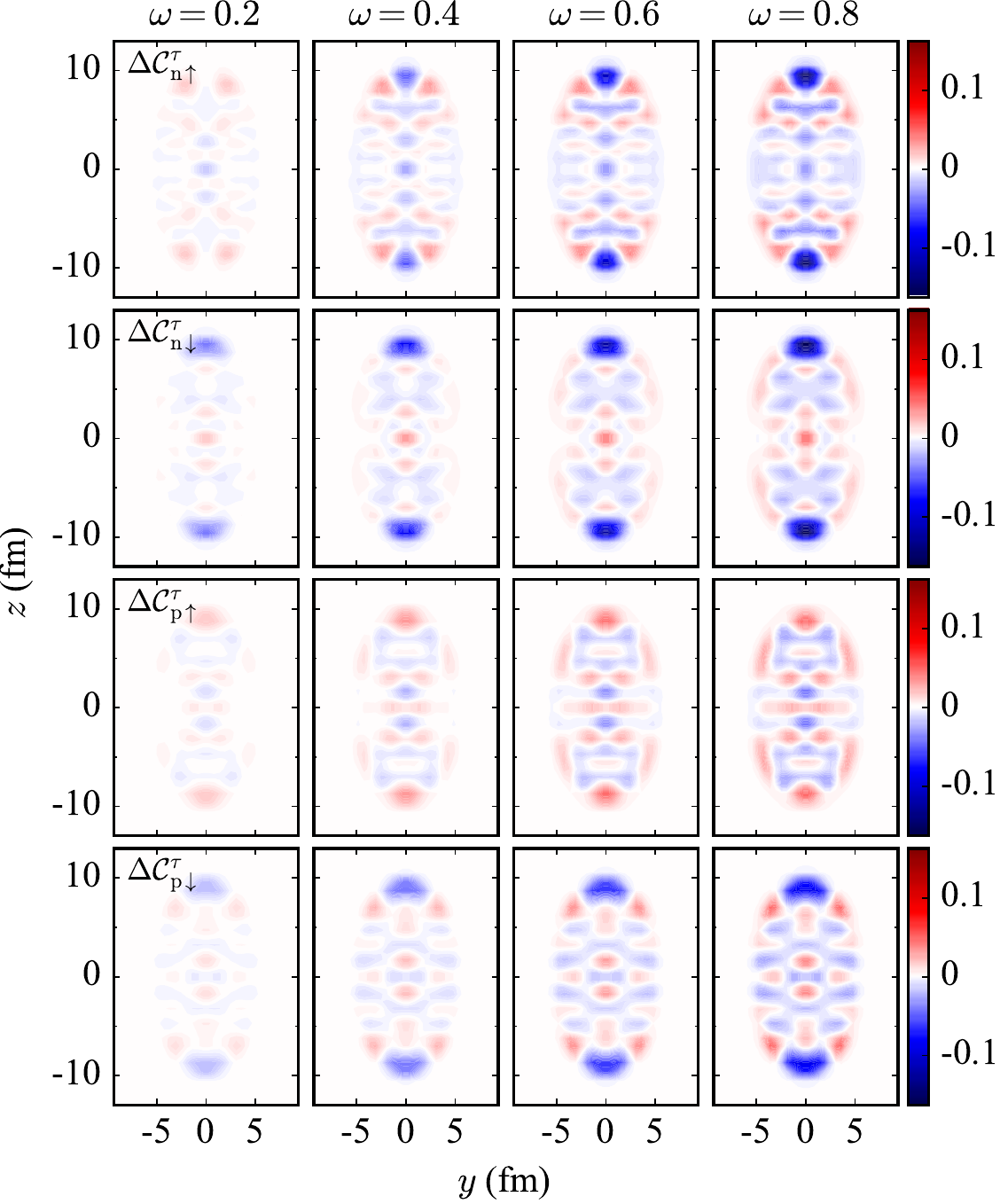}
	\caption{Similar to Fig.~\ref{fig:dy152_NLFtau_diff_xz} but for $\Delta{\cal C}_{q\sms}^\tau$  in  the  $y$-$z$ ($x=0$)  plane}
	\label{fig:dy152_NLFtau_diff_yz}
\end{figure}

Also, to illustrate the directional dependence of $\Delta{\cal C}_{q\sms}^\tau$, we show it in the  $x=0$  plane in
Fig.~\ref{fig:dy152_NLFtau_diff_yz}. A different pattern along the $y$ direction results from the breaking of axial symmetry by the  cranking term.

\begin{figure}[htb]
	\includegraphics[width=\linewidth]{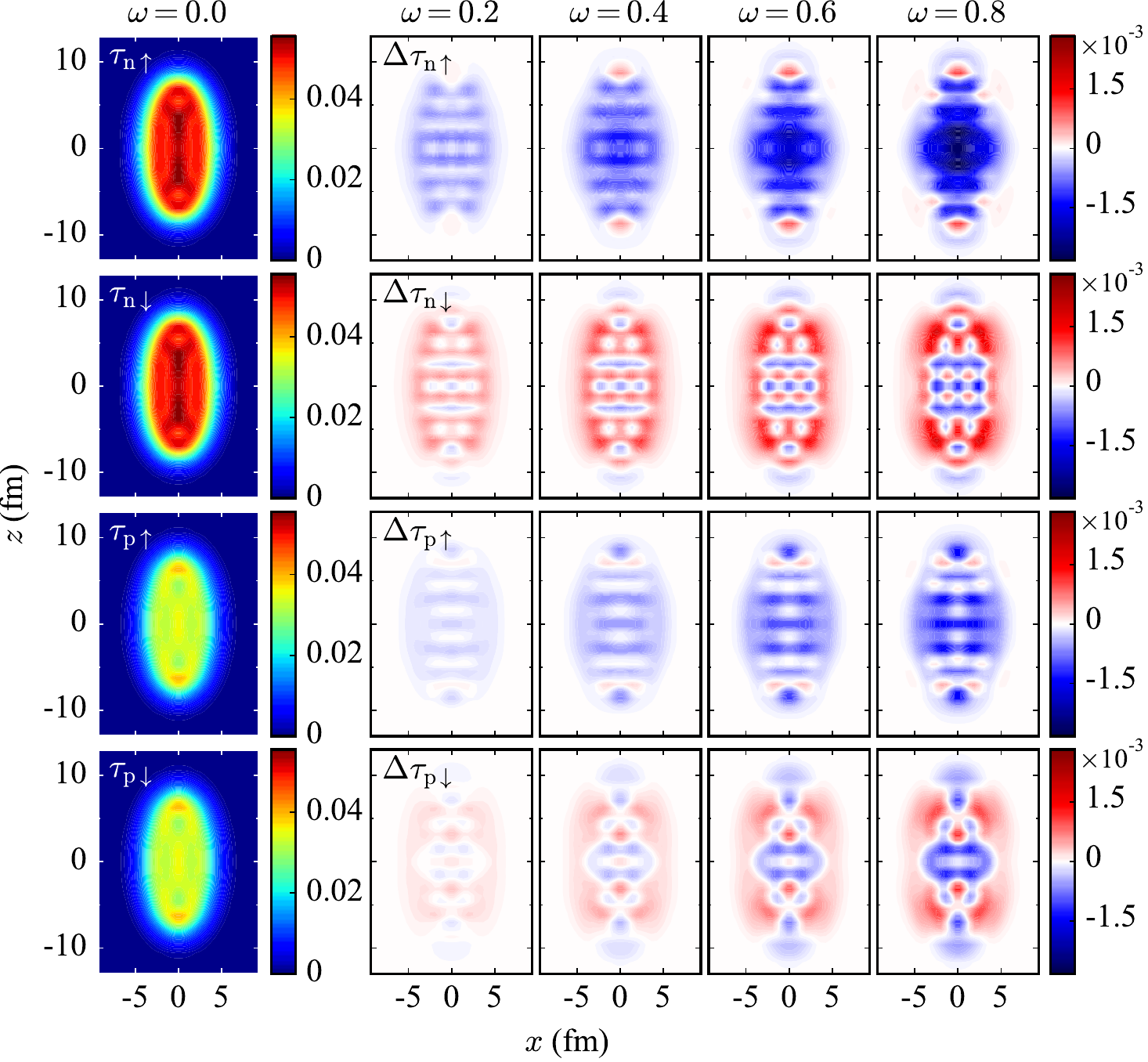}
	\caption{Similar to Fig.\ \ref{fig:dy152_NLFtau_diff_xz}, but for  $\Delta\tau_{q\sms}$ (in fm$^{-5}$). The reference value of $\tau_{q\sms}$ at $\omega=0$ is shown in the first column.}
	\label{fig:dy152_tau_diff_xz}
\end{figure}
\begin{figure}[htb]
	\includegraphics[width=\linewidth]{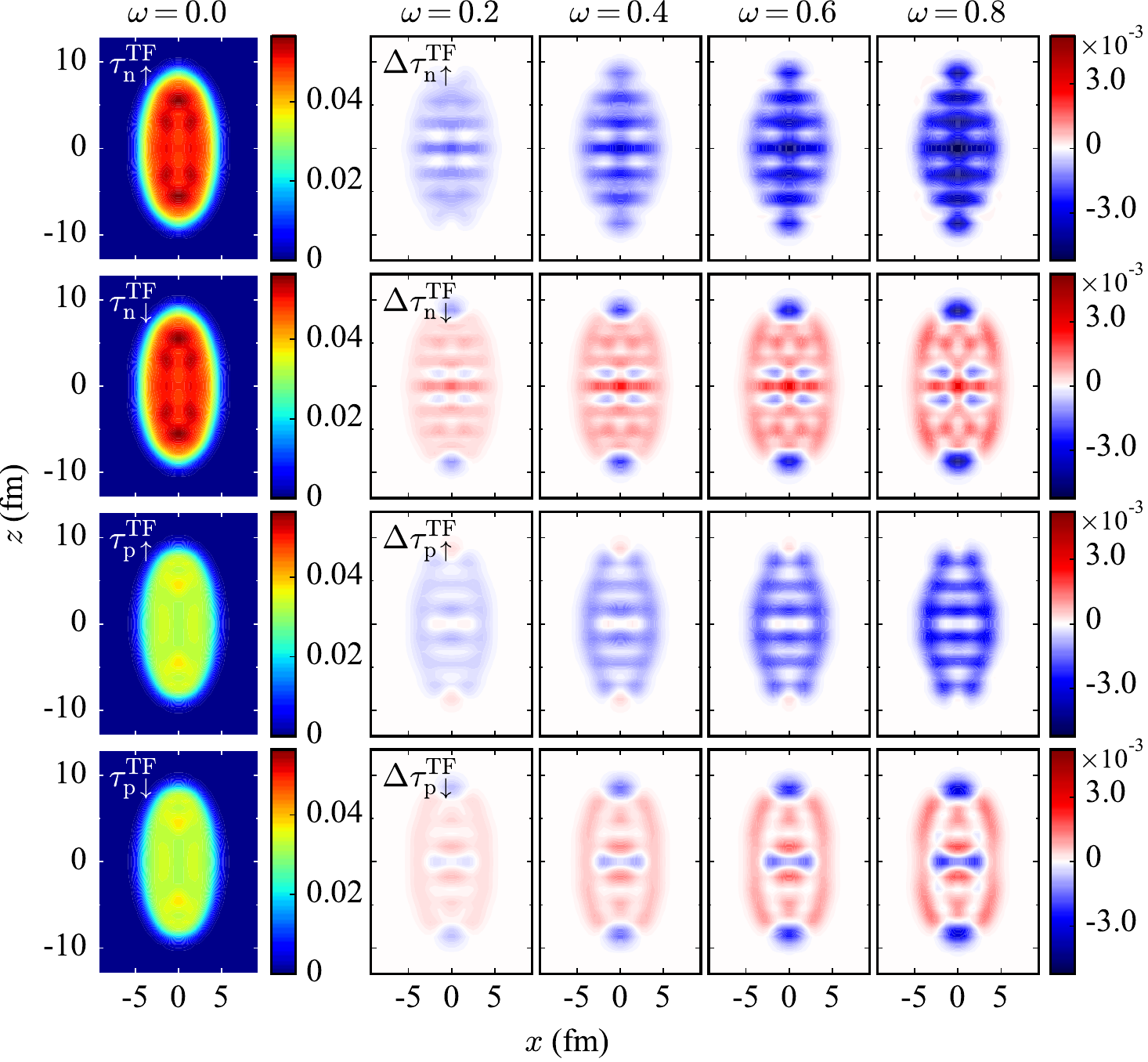}
	\caption{Similar to Fig.\ \ref{fig:dy152_tau_diff_xz}, but for 
	$\Delta\tau_{q\sms}^{\mathrm{TF}}$ (in fm$^{-5}$).}
	\label{fig:dy152_tauTF_diff_xz}
\end{figure}
Figures \ref{fig:dy152_tau_diff_xz} and \ref{fig:dy152_tauTF_diff_xz} show the variations of $\Delta\tau$ and $\Delta\tau^{\mathrm{TF}}$ with $\omega$.
Similar to the CHO case discussed in Sec.~\ref{sec:results_ho},
$\Delta \tau$ and  $\Delta \tau^{\mathrm{TF}}$ are in antiphase that results in a constructive interference when it comes to  $\Delta \mathcal{C}^\tau$. \
Furthermore, $\Delta \tau$, $\Delta \tau^{\mathrm{TF}}$, and $\Delta \mathcal{C}_{q\sms}$ of opposite values of $\sms$  change in the opposite direction with  $\omega$. 
That is, a ridge in $\Delta\mathcal{C}_{q\uparrow}$ corresponds to a valley in $\Delta\mathcal{C}_{q\downarrow}$. 
According to Eqs.~(\ref{eq:density_relation_simplex}), this is due to the contributions from time-odd densities which change sign between different values of $\sms$. 

By investigating  the behavior of the NLF from the perspective of individual  s.p.\ orbits, we can gain useful insights on the s.p.\ motion in the rotating nucleus. Following the CHO discussion in Sec.~\ref{sec:results_ho}, we
 focus on $\Delta \tau$. In particular, we  shall study the rotational dependence of kinetic-energy densities of s.p.\ orbits near the Fermi level as these orbits are expected
to primarily affect the nuclear response to rotation. 

In the  example discussed below,  for the sake of simplicity we consider  small rotational frequency $\hbar\omega = 0.1$\,MeV, at which individual levels shown in Fig.~\ref{fig:dy152_levels}  can easily be identified structurally. At higher rotational frequencies, this discussion can be
repeated by following the diabatic Routhians within each parity-signature block.
\begin{figure}[htb]
	\includegraphics[width=\linewidth]{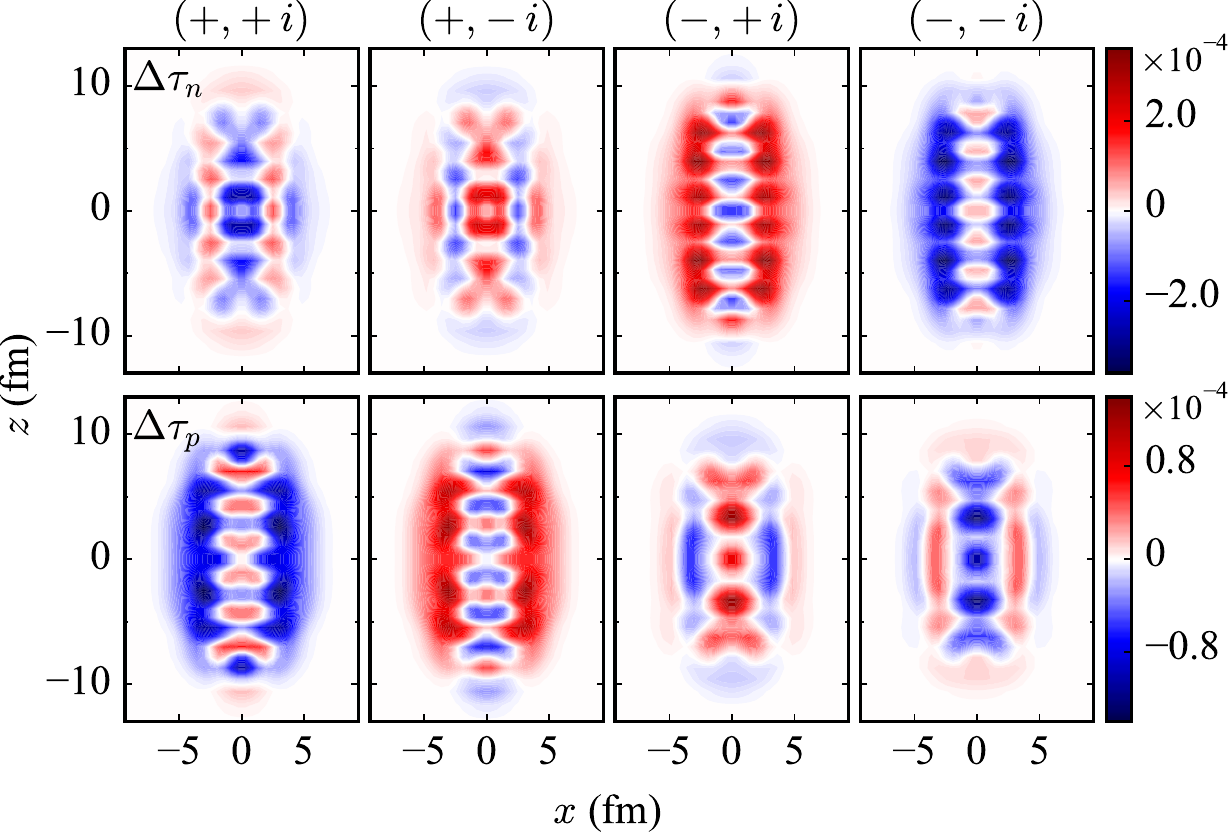}
	\caption{Neutron (top) and proton (bottom) contributions to $\Delta\tau$ (in fm$^{-5}$) in the $x$-$z$ ($y=0$) plane for different parity-signature blocks ($\pi, r_y$) in $^{152}$Dy at $\hbar\omega=0.1$\,MeV.}
	\label{fig:dy152_tau_diff_paritysig_xz}
\end{figure}

\begin{figure}[htb]
	\includegraphics[width=\linewidth]{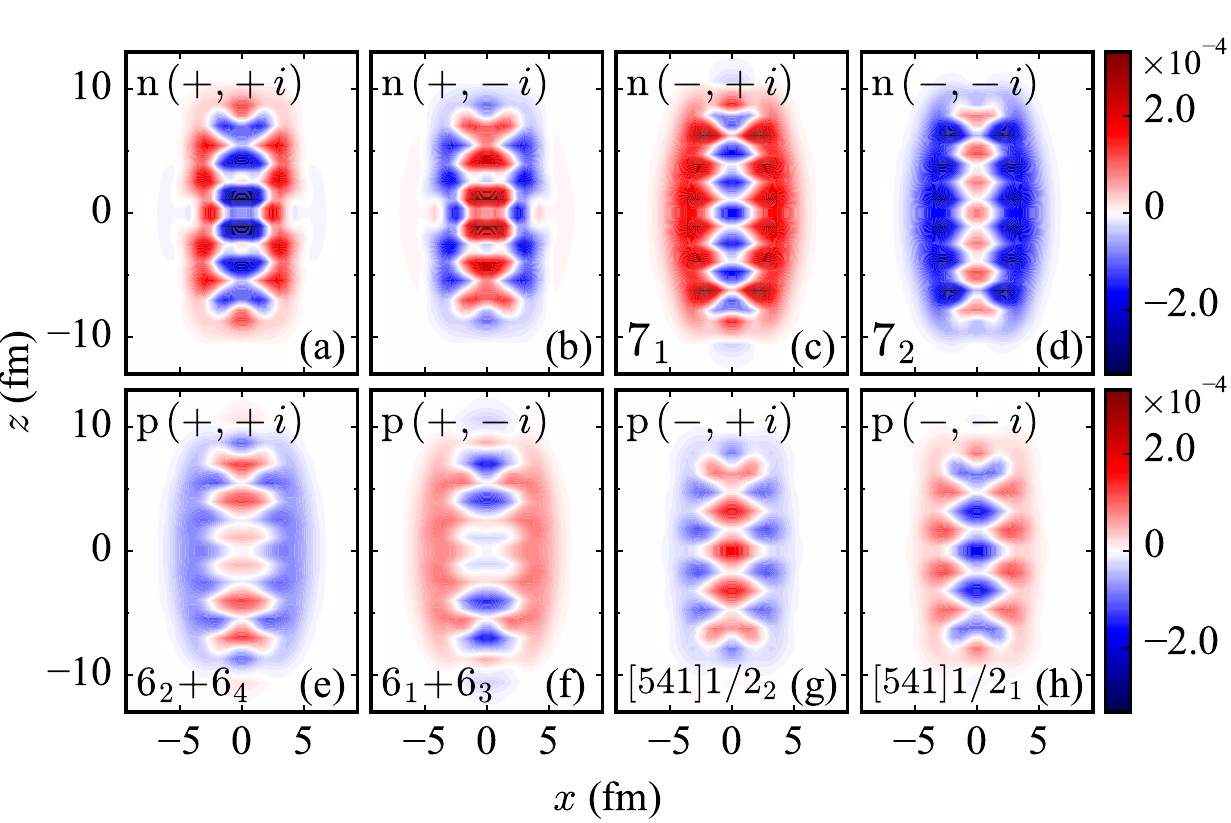}
	\caption{Contributions to $\Delta\tau$ (in fm$^{-5}$) in the $x$-$z$ ($y=0$) plane for different parity-signature blocks from individual s.p.\ Routhians  in $^{152}$Dy at  $\hbar\omega=0.1$\,MeV: the four $\pi=+, r=+i$ neutron levels 
 [651]1/2, [642]5/2, [413]5/2, and [411]1/2 with (a) $r=+i$ and (b) $-i$ that appear  below the  $N=86$ shell gap in Fig.\ \ref{fig:dy152_levels} (see Fig.\ 1 of Ref.\ \cite{Nazarewicz1989} for the  asymptotic (Nilsson) quantum numbers $[{\cal N}n_z \Lambda]\Omega$ of s.p.\ levels in SD $^{152}$Dy); the ${\cal N}=7$ neutron intruder states  (c) 7$_1$ and (d) 7$_2$;
 the ${\cal N}=6$ proton intruder states (e) $6_2+6_4$ and (f) $6_1+6_3$;  and the (g) [541]1/2$_2$
  and (h) [541]1/2$_1$ proton states.
}
	\label{fig:dy152_sp_contribution_xz}
\end{figure}

Figure~\ref{fig:dy152_tau_diff_paritysig_xz} shows $\Delta\tau$ at $\hbar\omega=0.1$\,MeV for  different parity-signature blocks.
The patterns of $\Delta\tau$  can be understood  by inspecting  the contributions from several individual s.p.\ orbits  close to the Fermi energy shown in Fig.\ \ref{fig:dy152_sp_contribution_xz}. The main contribution to $\Delta\tau_n$ in the negative-parity blocks comes from the high-${\cal N}$ orbits $7_1$ and $7_2$. For the $\pi=+$ neutrons, four close-lying deformation-aligned states [651]1/2, [642]5/2, [413]5/2, and [411]1/2, are most important.
For the protons, the main contributions to $\Delta\tau$ come from
the ${\cal N}=6$ states  $6_1, 6_2, 6_3$, and $6_4$ (for $\pi=+$) and [541]1/2
(for $\pi=-$). 		 
It is seen that the s.p.\ contributions shown in  Fig.~\ref{fig:dy152_sp_contribution_xz}(a-h)  explain the behavior of
$\Delta\tau$  in  Fig.\ \ref{fig:dy152_tau_diff_paritysig_xz}.  
As discussed earlier in Sec.\ \ref{sec:results_ho}, characteristic nodal structures of $\Delta \tau$ along the $z$ axis primarily come from the evolution of rotation-aligned s.p.\ orbits with large ${\cal N}$ and $n_z$,  below the Fermi energy. The  features in the direction of the minor axis  can be  attributed to deformation-aligned s.p.\ states.

\section{Conclusions}\label{sec:conclusion}

In this study, we extended the concept of the fermion localization function to anisotropic, spin-unsaturated, and spin-polarized systems. In particular, we considered the case of broken time-reversal symmetry. We demonstrated that, in the general case of rotating deformed systems,  three localization measures
$\mathcal{C}_{qs_\mu}(\boldsymbol{r})$, with $\mu=x,y,z$, which depend on the anisotropy of the spin distribution, can be defined. 

We used the NLF to interpret  the results of cranked Skyrme-HF calculations for rotating nuclei, especially  to study the interplay between collective  and s.p.\ motion.
While the standard probabilistic interpretation  of the NLF cannot be easily extended to the case of self-consistent symmetries associated with point groups, such as signature or simplex, there are no conceptual problems  when viewing  the NLF as a measure of the excess of kinetic-energy density due to the Pauli principle.

The localization function involves various  local densities, among which the current density $\boldsymbol{j}$, density gradient $\vec{\nabla}\rho$, and  spin-current tensor density $\mathbb{J}$ are appreciable only  in the surface region. If one
neglects these surface terms,  one can define  a simplified localization measure ${\cal C}^\tau$, which involves only the kinetic-energy density $\tau$ and the Thomas-Fermi kinetic energy density $\tau^{\mathrm{TF}}$. We argue that ${\cal C}^\tau$ is amplified by the out-of-phase spatial oscillation of $\tau$ and $\tau^{\mathrm{TF}}$ attributed to the specific nodal structure of high-${\cal N}$ s.p.\ states.

To show the usefulness of the extended NLF, we carried out the Skyrme-CHF analysis of the  superdeformed yrast band of  $^{152}$Dy.
As the rotational frequency increases, 
rotationally aligned s.p.\ states with high-${\cal N}$ and  high-$n_z$  produce  a characteristic  oscillating pattern  in the NLF along the major axis of the nucleus, while the pattern variations along the minor axis  come from deformation-aligned s.p.\ states close to the Fermi energy. 

Our CHF and CHO results demonstrate that ${\cal C}^\tau$ is an excellent indicator of the nuclear response to collective rotation. Many applications of the NLF to the  visualization of nuclear  rotational and vibrational modes and time dependent processes \cite{Dumitrescu1984,Iwata2011,Xu2013,Xia2014,schuetrumpfClusterFormationPrecompound2017,Godbey2017,petrikThoulessValatinRotationalMoment2018,Nesterenko2018} are envisioned, especially after incorporating pairing correlations  via the HFB extension of the formalism.
One can also consider applying  the concept of the NLF beyond the mean-field approach. In particular, since the  kinetic-energy density can be computed within realistic $A$-body frameworks \cite{gennariNuclearKineticDensity2019}, studies of many-body correlations with  the help of ${\cal C}^\tau$ could offer new perspectives.

Finally, let us note that, while in the usual atomic applications the current term in Eq.~(\ref{eq:prob_D}) is ignored, the contribution to ELF from the spin-current tensor density $\mathbb{J}$ is expected to be nonzero in relativistic superheavy atoms. For instance, the spin-orbit splitting for the valence $7p$ orbital of  the element Og ($Z=118$) is predicted to be very large, around 10\,eV \cite{jerabekElectronNucleonLocalization2018,Schwerdtfeger2020}. While Og is believed to be a spin-saturated system (the whole 7$p$ shell is filled), this is not the case for, e.g.,  Fl ($Z=114$, $7p_{3/2}$ shell empty) for which $\mathbb{J}$ and the resulting spin-orbit current should be consider when analyzing the corresponding ELF.

\begin{acknowledgements}
Discussions with Jacek Dobaczewski, Samuel Giuliani,  and Simin Wang are gratefully acknowledged.
Computational resources were provided by the Institute for Cyber-Enabled Research at Michigan State University. 
This material is based upon work supported by the U.S.\ Department of Energy, Office of Science, Office of Nuclear Physics under award numbers DE-SC0013365 and DE-SC0018083 (NUCLEI SciDAC-4 Collaboration).
\end{acknowledgements}

%

\end{document}